\documentclass[twocolumn,twocolappendix]{aastex63}

\usepackage{natbib,aas_macros,amsmath}
\usepackage{multirow,color}
\usepackage{natbib}

\newcommand{\sii}{[S\,{\sc ii}]}
\newcommand{\oiii}{[O\,{\sc iii}]}
\newcommand{\oii}{[O\,{\sc ii}]}
\newcommand{\cii}{[C\,{\sc ii}]}

\newcommand{\ciii}{C\,{\sc iii}]}

\newcommand{\hst}{{\it HST}}
\newcommand{\jwst}{{\it JWST}}
\newcommand{\spitzer}{{\it Spitzer}}
\received{December 15, 2022}
\revised{December 13, 2023}
\accepted{\today}
\shorttitle{
JWST and ALMA Study on a Galaxy at $z=8.496$
}
\shortauthors{Fujimoto et al.}

\begin{document}

\title{
JWST and ALMA Multiple-Line Study in and around a Galaxy at \boldmath $z=8.496$:\\
Optical to FIR Line Ratios and the Onset of an Outflow Promoting Ionizing Photon Escape
}

\correspondingauthor{Seiji Fujimoto}
\email{fujimoto@utexas.edu}
\author[0000-0001-7201-5066]{Seiji Fujimoto}\altaffiliation{Hubble Fellow}
\affiliation{Department of Astronomy, The University of Texas at Austin, Austin, TX 78712, USA}
\affiliation{Cosmic Dawn Center (DAWN), Denmark}
\affiliation{Niels Bohr Institute, University of Copenhagen, Lyngbyvej 2, DK2100 Copenhagen \O, Denmark}


\author[0000-0002-1049-6658]{Masami Ouchi}
\affiliation{National Astronomical Observatory of Japan, 2-21-1 Osawa, Mitaka, Tokyo 181-8588, Japan}
\affiliation{Institute for Cosmic Ray Research, The University of Tokyo, 5-1-5 Kashiwanoha, Kashiwa, Chiba 277-8582, Japan}
\affiliation{Kavli Institute for the Physics and Mathematics of the Universe (WPI), University of Tokyo, Kashiwa, Chiba 277-8583, Japan}

\author[0000-0003-2965-5070]{Kimihiko Nakajima}
\affiliation{National Astronomical Observatory of Japan, 2-21-1 Osawa, Mitaka, Tokyo 181-8588, Japan}

\author[0000-0002-6047-430X]{Yuichi Harikane}
\affiliation{Institute for Cosmic Ray Research, The University of Tokyo, 5-1-5 Kashiwanoha, Kashiwa, Chiba 277-8582, Japan}

\author[0000-0001-7730-8634]{Yuki Isobe}
\affiliation{Institute for Cosmic Ray Research, The University of Tokyo, 5-1-5 Kashiwanoha, Kashiwa, Chiba 277-8582, Japan}
\affiliation{Department of Physics, Graduate School of Science, The University of Tokyo, 7-3-1 Hongo, Bunkyo, Tokyo 113-0033, Japan}


\author[0000-0003-2680-005X]{Gabriel Brammer}
\affiliation{Cosmic Dawn Center (DAWN), Denmark}
\affiliation{Niels Bohr Institute, University of Copenhagen, Lyngbyvej 2, DK2100 Copenhagen \O, Denmark}


\author[0000-0003-3484-399X]{Masamune Oguri}
\affiliation{Center for Frontier Science, Chiba University, 1-33 Yayoi-cho, Inage-ku, Chiba 263-8522, Japan}
\affiliation{Department of Physics, Graduate School of Science, Chiba University, 1-33 Yayoi-Cho, Inage-Ku, Chiba 263-8522, Japan}


\author[0000-0001-9419-9505]{Clara~Gim\'enez-Arteaga}
\affiliation{Cosmic Dawn Center (DAWN), Denmark}
\affiliation{Niels Bohr Institute, University of Copenhagen, Blegdamsvej 17, DK2100 Copenhagen \O, Denmark}

\author[0000-0002-9389-7413]{Kasper~E.~Heintz}
\affiliation{Cosmic Dawn Center (DAWN), Denmark}
\affiliation{Niels Bohr Institute, University of Copenhagen, Jagtvej 128, DK-2200 Copenhagen N, Denmark}

\author[0000-0002-5588-9156]{Vasily Kokorev}
\affiliation{Cosmic Dawn Center (DAWN), Denmark}
\affiliation{Niels Bohr Institute, University of Copenhagen, Blegdamsvej 17, DK2100 Copenhagen \O, Denmark}


\author[0000-0002-8686-8737]{Franz E. Bauer}
\affiliation{Instituto de Astrof{\'{\i}}sica, Facultad de F{\'{i}}sica, Pontificia Universidad Cat{\'{o}}lica de Chile, Campus San Joaquín, Av. Vicuña Mackenna 4860, Macul Santiago, Chile, 7820436} 
\affiliation{Centro de Astroingenier{\'{\i}}a, Facultad de F{\'{i}}sica, Pontificia Universidad Cat{\'{o}}lica de Chile, Campus San Joaquín, Av. Vicuña Mackenna 4860, Macul Santiago, Chile, 7820436} 
\affiliation{Millennium Institute of Astrophysics, Nuncio Monse{\~{n}}or S{\'{o}}tero Sanz 100, Of 104, Providencia, Santiago, Chile} 

\author[0000-0002-9400-7312]{Andrea Ferrara}
\affiliation{
Scuola Normale Superiore, Piazza dei Cavalieri 7, 50126 Pisa, Italy
}

\author[0000-0001-5780-1886]{Takashi Kojima}
\affiliation{Institute for Cosmic Ray Research, The University of Tokyo, 5-1-5 Kashiwanoha, Kashiwa, Chiba 277-8582, Japan}

\author[0000-0003-3021-8564]{Claudia del P. Lagos}
\affiliation{International Centre for Radio Astronomy Research (ICRAR), M468, University of Western Australia, 35 Stirling Hwy, Crawley, \\WA 6009, Australia.}
\affiliation{ARC Centre of Excellence for All Sky Astrophysics in 3 Dimensions (ASTRO 3D).}
\affiliation{Cosmic Dawn Center (DAWN), Denmark}

\author[0000-0002-2906-2200]{Sommovigo Laura}
\affiliation{Scuola Normale Superiore, Piazza dei Cavalieri 7, 50126 Pisa, Italy}

\author[0000-0001-7144-7182]{Daniel Schaerer}
\affiliation{Observatoire de Genève, Université de Genève, Chemin Pegasi 51, 1290 Versoix, Switzerland}
\affiliation{CNRS, IRAP, 14 Avenue E. Belin, 31400 Toulouse, France}

\author[0000-0002-2597-2231]{Kazuhiro Shimasaku}
\affiliation{Department of Astronomy, School of Science, The
University of Tokyo, 7-3-1 Hongo, Bunkyo-ku, Tokyo 113-0033, Japan}
\affiliation{Research Center for the Early Universe, The University of
Tokyo, 7-3-1 Hongo, Bunkyo-ku, Tokyo 113-0033, Japan}


\author[0000-0001-6469-8725]{Bunyo Hatsukade}
\affiliation{Institute of Astronomy, Graduate School of Science, The University of Tokyo, 2-21-1 Osawa, Mitaka, Tokyo 181-0015, Japan}

\author[0000-0002-4052-2394]{Kotaro Kohno}
\affiliation{Institute of Astronomy, Graduate School of Science, The University of Tokyo, 2-21-1 Osawa, Mitaka, Tokyo 181-0015, Japan}
\affiliation{Research Center for the Early Universe, Graduate School of Science, The University of Tokyo, 7-3-1 Hongo, Bunkyo-ku, Tokyo 113-0033, Japan}

\author[0000-0002-4622-6617]{Fengwu Sun}
\affiliation{Steward Observatory, University of Arizona, 933 N. Cherry Avenue, Tucson, AZ 85721, USA}

\author[0000-0001-6477-4011]{Francesco Valentino}
\affiliation{Cosmic Dawn Center (DAWN), Denmark}
\affiliation{Niels Bohr Institute, University of Copenhagen, Jagtvej 128, DK-2200 Copenhagen N, Denmark}
\affiliation{European Southern Observatory, Karl-Schwarzschild-Str. 2, D-85748 Garching bei Munchen, Germany}

\author[0000-0002-4465-8264]{Darach Watson}
\affiliation{Cosmic Dawn Center (DAWN), Denmark}
\affiliation{Niels Bohr Institute, University of Copenhagen, Jagtvej 128, DK-2200 Copenhagen N, Denmark}


\author[0000-0001-7440-8832]{Yoshinobu Fudamoto}
\affiliation{
Waseda Research Institute for Science and Engineering, Faculty of Science and Engineering, Waseda University, 3-4-1 Okubo, Shinjuku, Tokyo 169-8555, Japan
}
\affiliation{
National Astronomical Observatory of Japan, 2-21-1, Osawa, Mitaka, Tokyo, Japan
}

\author[0000-0002-7779-8677]{Akio K. Inoue}
\affiliation{
Waseda Research Institute for Science and Engineering, Faculty of Science and Engineering, Waseda University, 3-4-1 Okubo, Shinjuku, Tokyo 169-8555, Japan
}
\affiliation{
Department of Physics, School of Advanced Science Engineering, Faculty of Science and Engineering, Waseda University, 3-4-1 Okubo, Shinjuku, Tokyo 169-8555, Japan
}

\author[0000-0003-3926-1411]{Jorge Gonz\'alez-L\'opez}
\affiliation{
N\'ucleo de Astronom\'ia de la Facultad de Ingenier\'ia y Ciencias, Universidad Diego Portales, Av. Ejército Libertador 441, Santiago, Chile
}
\affiliation{
Las Campanas Observatory, Carnegie Institution of Washington, Casilla 601, La Serena, Chile
}

\author[0000-0002-6610-2048]{Anton M. Koekemoer}
\affiliation{Space Telescope Science Institute, 3700 San Martin Dr., Baltimore, MD 21218, USA}

\author[0000-0002-7821-8873]{Kirsten Knudsen}
\affiliation{
Department of Space, Earth and Environment, Chalmers University of Technology, Onsala Space Observatory, SE-43992 Onsala, Sweden
}

\author[0000-0002-2419-3068]{Minju~M.~Lee}
\affiliation{Cosmic Dawn Center (DAWN), Denmark} 
\affiliation{DTU-Space, Technical University of Denmark, Elektrovej 327, DK2800 Kgs. Lyngby, Denmark}

\author[0000-0002-4872-2294]{Georgios E. Magdis}
\affiliation{Cosmic Dawn Center (DAWN), Jagtvej 128, DK2200 Copenhagen N, Denmark}
\affiliation{DTU-Space, Technical University of Denmark, Elektrovej 327, 2800, Kgs. Lyngby, Denmark}
\affiliation{Niels Bohr Institute, University of Copenhagen, Jagtvej 128, 2200, Copenhagen N, Denmark}

\author[0000-0001-5492-1049]{Johan Richard}
\affiliation{
Univ Lyon, Univ Lyon1, Ens de Lyon, CNRS, Centre de Recherche Astrophysique de Lyon UMR5574, F-69230, Saint-Genis-Laval,France
}

\author[0000-0002-6338-7295]{Victoria~B.~Strait}
\affiliation{Cosmic Dawn Center (DAWN), Denmark}
\affiliation{Niels Bohr Institute, University of Copenhagen, Jagtvej 128, DK-2200 Copenhagen N, Denmark}

\author[0000-0001-6958-7856]{Yuma Sugahara}
\affiliation{Waseda Research Institute for Science and Engineering, Faculty of Science and Engineering, Waseda University, 3-4-1 Okubo, Shinjuku, Tokyo 169-8555, Japan}
\affiliation{National Astronomical Observatory of Japan, 2-21-1, Osawa, Mitaka, Tokyo, Japan}

\author[0000-0003-4807-8117]{Yoichi Tamura}
\affiliation{Division of Particle and Astrophysical Science, Graduate School of Science, Nagoya University, Nagoya 464-8602, Japan}

\author[0000-0003-3631-7176]{Sune Toft}
\affiliation{Cosmic Dawn Center (DAWN), Denmark}
\affiliation{Niels Bohr Institute, University of Copenhagen, Jagtvej 128, DK-2200 Copenhagen N, Denmark}

\author[0000-0003-1937-0573]{Hideki Umehata}
\affiliation{Institute for Advanced Research, Nagoya University, Furocho, Chikusa,
Nagoya 464-8602, Japan}
\affiliation{Department of physics, Graduate School of Science, Nagoya University,
Nagoya, Aichi 464-8602, Japan}

\author[0000-0002-6313-6808]{Gregory Walth}
\affiliation{IPAC, California Institute of Technology, Mail Code 314-6, 1200 E.
California Blvd., Pasadena, CA 91125}

\def\apj{ApJ}%
\def\apjl{ApJ}%
\def\apjs{ApJS}%

\def\rme{\rm e}
\def\rmstar{\rm star}
\def\rmFIR{\rm FIR}
\def\itHubble{\it Hubble}
\def\rmyr{\rm yr}
\def\targ{ID4590}

\begin{abstract} 
We present ALMA deep spectroscopy for a lensed galaxy at $z_{\rm spec}=8.496$ with $\log(M_{\rm star}/M_{\odot})\sim7.8$ whose optical nebular lines and stellar continuum are detected by \jwst/NIRSpec and NIRCam Early Release Observations in SMACS0723. Our ALMA spectrum shows \oiii88~$\mu$m and \cii158~$\mu$m line detections at $4.0\sigma$ and $4.5\sigma$, respectively.
The redshift and position of the \oiii\ line coincide with those of the \jwst\ source, while the \cii\ line is blue-shifted by 90~km~s$^{-1}$ with a spatial offset of $0\farcs5$ ($\approx0.5$~kpc in source plane) from the centroid of the \jwst\ source. The NIRCam F444W image, including \oiii$\lambda$5007 and H$\beta$ line emission, spatially extends beyond the stellar components by a factor of $>8$.  
This indicates that the $z=8.5$ galaxy has already experienced strong outflows 
as traced by extended \oiii$¥lambda$5007 and offset \cii\ emission, 
which would promote ionizing photon escape and facilitate reionization. 
With careful slit-loss corrections and 
removals of emission spatially outside the galaxy,
we evaluate the \oiii88$\mu$m/$\lambda$5007 line ratio, and derive the electron density $n_{\rm e}$ by photoionization modeling to be $220^{+230}_{-130}$~cm$^{-3}$, which is comparable with those of $z\sim2$--3 galaxies. We estimate an \oiii88$\mu$m/\cii158$\mu$m line ratio in the galaxy of $>4$, 
as high as those of known $z\sim$6--9 galaxies. 
This high \oiii88~$\mu$m/\cii158~$\mu$m line ratio is generally explained by the high $n_{\rm e}$ as well as the low metallicity ($Z_{\rm gas}/Z_{\odot}=0.04^{+0.02}_{-0.02}$), high ionization parameter ($\log U > -2.27$), and low carbon-to-oxygen abundance ratio ($\log$(C/O) $=[-0.52:-0.24]$) obtained from the \jwst/NIRSpec data; 
further \cii\ follow-up observations will constrain the covering fraction of photodissociation regions. 
\end{abstract}
\keywords{ galaxies: formation --- galaxies: evolution --- galaxies: high-redshift }

\section{Introduction}\label{sec:intro}
Studying early systems in the Epoch of Reionization (EoR; $z\gtrsim6$) is key to understanding fundamental cosmological questions such as the development of large-scale structure, the processes of cosmic reionization, and the first galaxy formation in the universe. 
In the last decades, deep {\it Hubble\,Space\,Telescope} ({\it HST}) surveys provided thousands of EoR galaxies and initial characterization of their stellar component, in terms of un-obscured star formation and sizes in the rest-frame ultra-violet (UV) wavelengths \citep[e.g.,][]{ellis2013, bouwens2015, finkelstein2015, oesch2016}. 

The Atacama Large Millimeter/submillimeter Array (ALMA) offers a unique rest-frame far-infrared (FIR) window on EoR galaxies to characterize the dust and gas properties of their interstellar media (ISMs), where the major cooling lines (e.g.,\ \cii\ 158$\mu$m, \oiii\ 88$\mu$m) and underlying dust continuum are probes of key mechanisms in the ISM such as disk rotation \citep[e.g.,][]{smit2018,rizzo2020}, gravitational instability \citep[e.g.,][]{tadaki2018}, formation of the bulge, disk, and spiral arms \citep[e.g.,][]{tsukui2021,lelli2021}, galaxy mergers \citep[e.g.,][]{lefevre2020}, outflows \citep[e.g.,][]{spilker2018}, and dust-obscured star-forming activities \citep[e.g.,][]{bowler2022, inami2022}, which strongly complements rest-UV studies with \hst.  
Since elements produced in stars are returned to the ISM, the metal gas properties traced by the fine-structure lines, with their different ionization potentials and critical densities, provide powerful probes on the star-formation history and related physical conditions of the ISM, such as the temperature, density, ionization, and metal enrichment \citep[e.g.,][]{maiolino2019}. 

Previous ALMA observations have also raised questions
about what causes large diversity in EoR galaxies and the difference from local galaxies. 
Early deep ALMA \cii\ spectroscopy of some EoR galaxies has shown low $L_{\rm [CII]}$/SFR values \citep[e.g.,][]{ota2014,knudsen2016}, suggesting that they have characteristic ISM conditions such as high ionization parameter, low metallicity, or low gas density \citep[e.g.,][]{ferrara2019}. 
However, recent ALMA observations detected dust continuum \citep[e.g.,][]{watson2015,inami2022} and bright \cii\ lines \citep[e.g.,][]{bouwens2022}, pointing to a $L_{\rm [CII]}$/SFR relation similar to that of local main-sequence galaxies \citep[e.g.,][]{schaerer2020} and even low-mass sub-$L^{\star}$ galaxies at $z\gtrsim6$ \citep[e.g.,][]{fujimoto2021,molyneux2022}. 
Moreover, while ALMA observations of EoR galaxies have also shown high $L_{\rm [OIII]}$/$L_{\rm [CII]}$ ratios \citep[e.g.,][]{inoue2016, harikane2020,carniani2020,witstok2022} that are comparable or even higher than local dwarf galaxies \citep[e.g.,][]{cormier2015}, the main driver for the high ratios is still under debate \citep[e.g.,][]{harikane2020,vallini2021,katz2022,witstok2022}.  
While promising, past \hst\ and ALMA results have yet to provide a clear picture of the ISM physics in EoR galaxies. 

To understand ISM physics, one of the key quantities is the gas-phase metallicity ($Z_{\rm gas}$), which is a direct probe of chemical enrichment and, thus, the evolutionary stage of galaxies.  The flux of each metal line is determined by the abundance of that element and its emissivity \citep{aller1984}. Therefore, we can accurately estimate the abundance once we measure the emissivity based on the electron temperature $T_{\rm e}$, i.e., the so-called \textit{direct method} \citep[e.g.,][]{pilyugin2005,andrews2013}. However, since the direct method requires the detection of very faint auroral lines (e.g., \oiii$\lambda$4363), this has until recently only been possible for a handful of sources up to $z\sim3$ \citep[e.g.,][]{christensen2012, kojima2017, sanders2020}.

Our observational landscape has been revolutionized by the advent of \jwst. 
As part of the Early Release observation (ERO) programs of \jwst, deep NIRSpec multi shutter array (MSA) observations have been performed towards 
the massive lensing cluster SMACSJ0723, successfully detecting  
multiple nebular emission lines in the rest-frame UV to optical wavelengths and determining spectroscopic redshifts for distant lensed galaxies out to $z=8.496$ \citep[e.g.,][]{carnall2022, schaerer2022, trump2022, curti2022}. 
Remarkably, the deep NIRSpec spectra also show the detection of the auroral line \oiii$\lambda$4363, enabling the first direct temperature $T_{\rm e}$ method estimates at such high redshifts \citep[e.g.,][]{schaerer2022,trump2022, curti2022}. 
With the powerful \jwst\ and ALMA combination, we are finally in a position to fully reveal how the elements of galaxies -- gas, stellar, and dust -- interplay with each other and what governs the growth of galaxies in the infant Universe. 

In this paper, 
we present ALMA deep follow-up of two major coolant lines, \cii\ 158$\mu$m and \oiii\ 88$\mu$m, and underlying continuum for the lensed galaxy at $z=8.496$, whose warm ISM properties are best characterized by the latest deep \jwst\ observations, including the \oiii$\lambda$4363 line. 
This is the first FIR characterization of an EoR galaxy with a robust metallicity measurement via the \textit{direct method}, setting the benchmark to understand and interpret previous results from ALMA EoR galaxy studies over the last decade and providing a timely, unique reference for future follow-up of EoR galaxies in the coming decade.

The structure of this paper is as follows. 
In Section 2, we describe the observations and the data processing of both \jwst\ and ALMA. 
Section 3 outlines the analyses related to the mass model for the lensing cluster, measurements of flux, size, and morphology, and the optical--mm SED fitting.  
In Section 4, we present the results from the multiple-line studies and discuss the physical origins of these results. 
A summary of this study is presented in Section 5. 
Throughout this paper, we assume a flat universe with 
$\Omega_{\rm m} =$ 0.3, 
$\Omega_\Lambda =$ 0.7, 
$H_0 =$ 70 km s$^{-1}$ Mpc$^{-1}$, 
and the Chabrier initial mass function \citep[IMF;][]{chabrier2003}.  
We adopt an angular scale of $1''=4.63$~kpc for the target redshift at $z=8.496$.
We take the cosmic microwave background (CMB) effect into account and correct the flux measurements at submm and mm bands, following the recipe presented by \cite{dacunha2013} (see also e.g., \citealt{pallottini2015, zhang2016, lagache2018}).
\section{Observations \& Data Processing} 
\label{sec:data}

\setlength{\tabcolsep}{3pt}
\begin{table*}
\caption{ALMA DDT Observation \& Data Properties for \targ\ at $z=8.496$ in SM0723}
\vspace{-0.8cm}
\label{tab:obs_summary}
\begin{center}
\begin{tabular}{cccccccccccc}
\hline 
\hline
Target line & Band & $\lambda_{\rm cent}$ & Obs Date & Baseline & $N_{\rm ant}$ &  $T_{\rm int}$ & PWV  & beam & $\sigma_{\rm line}$    & $\sigma_{\rm cont}$  \\
 &   & ($\mu$m)  & (YYYY-MM-DD) & (m)     & & (min)  &  (mm)        &  ($''\times''$)    &  ($\mu$Jy/beam)    &  ($\mu$Jy/beam)          \\ 
(1) & (2)  & (3)  & (4) & (5)     & (6) & (7)  &  (8)  &  (9)    &  (10)    &  (11)          \\  \hline
\cii\ 158~$\mu$m & 5  & 1540 & 2022-10-17 & 15.1--629.3 (C3) & 44  &  85.7   & 0.5 & $1.35\times1.25$ & 225 & 11.6   \\
\oiii\ 88~$\mu$m & 7  & 854  & 2022-10-14 & 15.1--629.3 (C3) & 44  &  99.8   & 0.4 & $0.71\times0.58$   & 420 &  20.9  \\ \hline \hline
\end{tabular}
\end{center}
\vspace{-0.2cm}
\tablecomments{
(1) FIR fine-structure atomic cooling line targeted in this program. 
(2) ALMA Band.
(3) Central wavelength corresponding to the central sky frequency in the observation. 
(4) Observation date.
(5) Baseline range. The parenthesis shows the configuration. 
(6) Number of antennae.
(7) On-source integration time in minutes. 
(8) Mean precipitable water vapor (PWV) during the observations.
(9) Synthesized beam size (FWHM) in the natural-weighted image. 
(10--11) Standard deviation of the pixels. For the cube, we show the value in the channel corresponding to the line frequency in the 20-km~s$^{-1}$ data cube. 
}
\end{table*}

\begin{figure*}
\includegraphics[trim=0cm 0cm 0cm 0cm, clip, angle=0,width=1\textwidth]{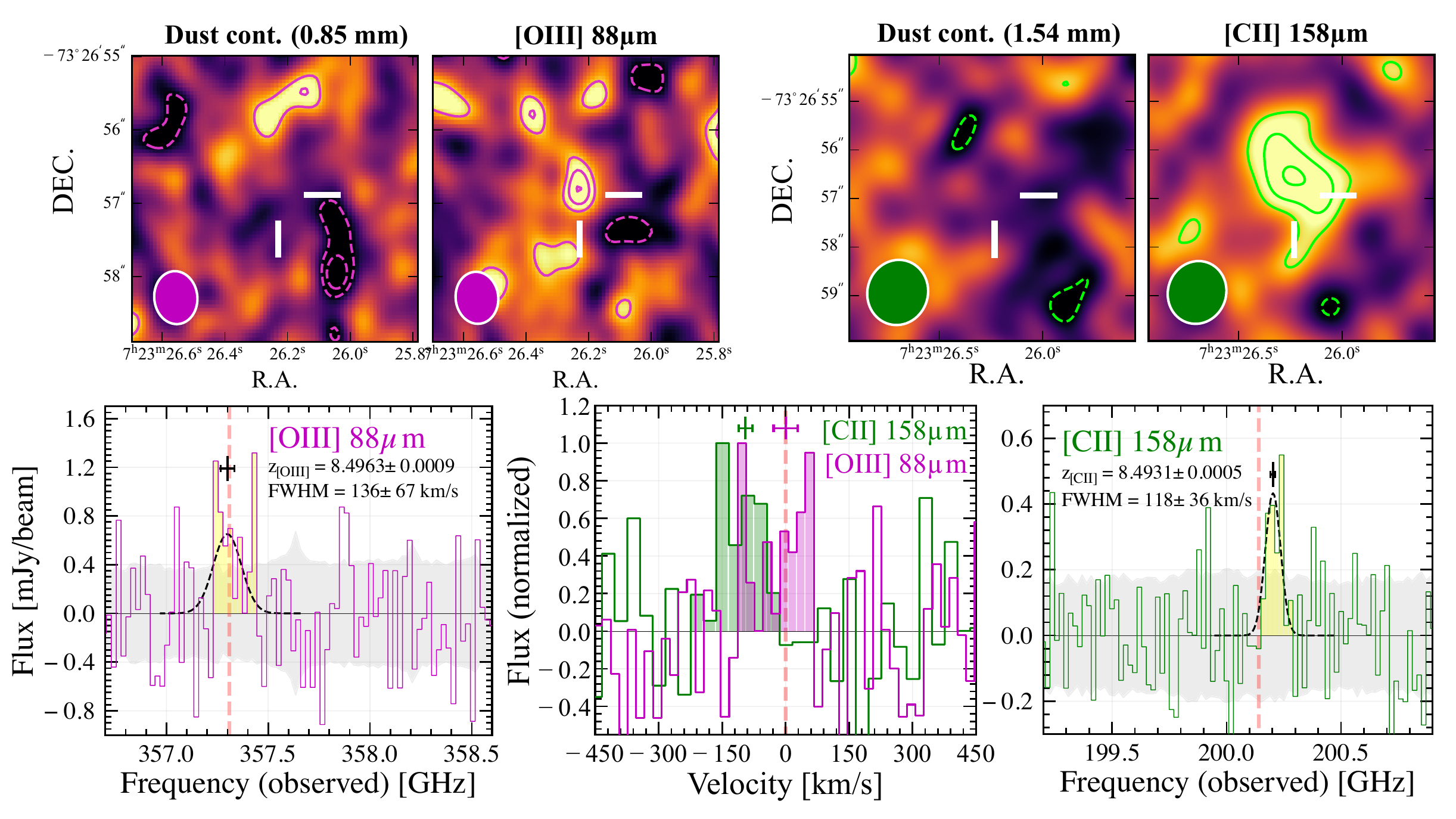}
 \caption{
Summary of ALMA DDT observation results. 
\textit{\textbf{Top:}} The ALMA dust continuum and velocity-integrated line maps for the \oiii88$\mu$m ($4''\times4''$) and \cii158$\mu$m lines ($6''\times6''$). 
The solid (dashed) contours are the 2$\sigma$, 3$\sigma$, and 4$\sigma$ (-3$\sigma$ and -2$\sigma$) levels.  
The white bars indicate the target position at the center of the maps, and the ellipse denotes the synthesized ALMA beam. 
\textit{\textbf{Bottom:}} 
ALMA line spectra, where the middle panel compares their line profiles in the velocity frame, where the zero velocity is based on $z=8.496$ determined by NIRSpec.
The spectra are extracted from the mean pixel count within optimized apertures smaller than the beam size at the source position. 
The dashed curve shows the best-fit Gaussian. 
The black bar shows the best-fit frequency center of the line with the 1$\sigma$ error, and the corresponding redshift and the line width are described in the label.  
The grey shade denotes the 1$\sigma$ error in each channel.
The red dashed vertical line denotes the expected line frequency based on the redshift of $z=8.496$ determined by NIRSpec.  
\label{fig:alma}}
\end{figure*}

\subsection{SMACS J0723.3-7327}
\label{sec:cluster}
The target field, SMACS J0723.3-7327 (hereafter SM0723), is a massive galaxy cluster at $z=0.390$ ($07^{h}23^{m}13.3^{s}$, $-73^{d}27^{m}25^{s}$), initially discovered via the Sunyaev-Zel'dovich effect in the Planck survey \citep{planck2011}. 
The galaxy cluster mass is estimated to be $M_{200} = 8.4\times10^{14}M_{\odot}$. 
SM0723 was observed as part of the \hst\ Reionization Lensing Cluster Survey (RELICS, \#GO-14096, PI: D. Coe, \citealt{coe2019}) treasury program and the subsequent \spitzer\ S-RELICS program (\#12005, PI: M. Bradac).

\subsection{ALMA}
\label{sec:alma_reduction}

ALMA Band~5 and Band~7 follow-up observations of \cii\ and \oiii\ spectroscopy were carried out on October 17th and 14th 2022, respectively, as a Cycle~8 DDT program (\#2022.A.00022.S, PI: S. Fujimoto).  
The target was a strongly lensed star-forming galaxy, \targ, spectroscopically confirmed at $z=8.496$ with the robust detection of rest-frame optical emission lines including the \oiii$\lambda$4363 line \citep[e.g.,][]{carnall2022, schaerer2022, trump2022, curti2022}. 
Both observations used the Frequency Division Mode, and baseline ranges 15--457~m in the C-3 configuration. 
The mean precipitable water vapour (PWV) was 0.5~mm and 0.4~mm, and the on-source integration times were 85.7~mins and 99.8~mins in Band~5 and Band~7, respectively. 
J0519-4546 was observed as a flux and bandpass calibrator in both observations, while  
the phase calibration was performed with J0601-7036 and J0635-7516 in Band~5 and Band~7 observations, respectively. 

The ALMA data were reduced and calibrated with the Common Astronomy Software Applications package version 6.4.1.12 (CASA; \citealt{casa2022}) with the pipeline script in the standard manner. 
We imaged the calibrated visibilities using natural weighting, a pixel scale of $0\farcs05$, and a primary beam limit down to 0.2 by running the CASA task {\sc tclean}. 
For cubes, 
we adopted spectral channel bins of 20, 30, and 40 km~s$^{-1}$ and performed the CLEAN algorithm. 
We confirmed the results unchanged much ($<$10~\%) in the following analyses via the choice of the spectral resolutions, and we used 20~km/s resolution cubes in the following analyses.
For continuum maps, 
the {\sc tclean} routines were executed down to the 1$\sigma$ level with a maximum iteration number of 100,000 in the automask mode \citep{kepley2020}. 
We mask the central $\pm120$~km~s$^{-1}$ channels to avoid contamination from the \cii\ and \oiii\ lines in the continuum map. 
The natural weighted images, which maximize the S/N, resulted in the full-width-half-maximum (FWHM) size of the synthesized beam of $1\farcs35\times1\farcs25$ and $0\farcs71\times0\farcs58$ with $1\sigma$ sensitivities for the continuum (line in a 20~km~s$^{-1}$ channel) of 11.6 (225) $\mu$Jy~beam$^{-1}$ and 20.9 (420) $\mu$Jy~beam$^{-1}$ in Band~5 and Band~7, respectively. 
Figure \ref{fig:alma} shows the reduced continuum maps and the line cubes for both ALMA Band~5 and Band~7. 
We summarize the data properties of the continuum maps and the line cubes in Table \ref{tab:obs_summary}. 

\subsection{JWST}
\label{sec:jwst}
As part of the \jwst\ ERO (\#ERO-2736; \citealt{pontoppidan2022}), SM0723 was observed with \jwst/NIRCam, NIRSpec, MIRI, and NIRISS in June 2022. 
In the NIRCam and MIRI observations, six (F090W, F150W, F200W, F277W, F356W, F444W) and four (F770W, F1000W, F1500W, F1800W) different filters were used with $\sim$7540~sec and $\sim$5600~sec exposure per filter, reaching 5$\sigma$ limiting magnitudes of $\sim$29.4--30.0 AB and $\sim$23--26.3 AB, respectively \citep{pontoppidan2022}.

For NIRCam and MIRI, 
we use publicly available reduced and calibrated imaging products via the \texttt{grizli} pipeline.\footnote{
\url{https://s3.amazonaws.com/grizli-v2/JwstMosaics/v4/index.html}} 
The detailed calibration and reduction procedures will be presented in Brammer et al. (in prep.) (see also, e.g., \citealt{bradley2022, fujimoto2022b}). 
Briefly, the \jwst\ pipeline calibrated level-2 NIRCam imaging products were retrieved and processed with the \texttt{grizli} pipeline \citep{brammer2021, brammer2023}, 
where the photometric zero-point correction was applied, 
including other corrections for ``snowballs''\footnote{
\url{https://jwst-docs.stsci.edu/data-artifacts-and-features/
snowballs-artifact}}, 
``wisps''\footnote{
\url{https://jwst-docs.stsci.edu/jwst-near-infrared-camera/
nircam-features-and-caveats/nircam-claws-and-wisps}}, 
and detector variations.\footnote{
\url{
https://github.com/gbrammer/grizli/pull/107}
} 
We include a systematic error of 10\% on the observed flux values in the following analyses as a conservative measure in the same manner as other recent NIRCam studies \citep[e.g.,][]{naidu2022, finkelstein2023}, 
while the derived photometric zeropoints are consistent with those derived by other teams for \jwst\ ERS programs \citep{boyer2022, nardiello2022}.
The fully-calibrated images in each filter were aligned with the GAIA DR3 catalog \citep{gaia2021}, co-added, and drizzled at a 20~mas and 40~mas pixel scale for the short-wavelength (SW: F090W, F150W, F200W) and long-wavelength (LW: F277W, F356W, F444W) NIRCam band images, respectively. 
For the produced maps, we also additionally correct the proper motion effects of the GAIA sources. 
In the Appendix, we show the residual astrometric offsets of the GAIA sources.
We confirm the residual astrometric offset with respect to the GAIA DR3 frame is close to zero with the uncertainty of $\sim$10--20 mas.
All the MIRI images (F770W, F1000W, F1500W, F1800W) are aligned, co-added, and drizzled at 40~mas pixel scale in the same manner.
Existing multi-wavelength WFC3 archival
imaging from \hst\ was also processed with \texttt{grizli}, being aligned, co-added, and drizzled at 40~mas pixel scale in the same manner (see also \citealt{kokorev2022}). 
We include all MIRI and \hst/F105W, F125W, F140W, and F160W data in our analysis. 
Given the existence of nearby objects (Section \ref{sec:morph}), we adopt $0\farcs36$-diameter aperture photometry which is corrected to the total flux measurement by \texttt{MAG\_AUTO}. 
We also correct for the Galactic dust reddening in the target direction. 
The \jwst\ and \hst\ photometry used in this paper are summarized in the Appendix. 

For NIRSpec, 
there are several studies applying the latest reduction and calibration (\citealt{heintz2022c}; \citealt{nakajima2023}), compared to previous studies \citep[e.g.,][]{schaerer2022, curti2022, trump2022, brinchmann2022, arellano2022}. 
In our paper, we use results from \cite{nakajima2023} due to requirements of specific parameter sets for the photoionization model analysis in Section \ref{sec:oiii-cii_ratio} that are self-consistently derived in \cite{nakajima2023}. 
We confirm that 
the measurements and the derived physical parameters are generally consistent with the previous studies within the errors \citep[e.g.,][]{schaerer2022, curti2022, trump2022, brinchmann2022, arellano2022} as well as the similarly latest calibration and reduction efforts presented in \cite{heintz2022c}. 
Briefly, the NIRSpec 1D spectra are re-created by using four of the six exposures after removing one with no signal and another with a noisy 2D spectrum around H$\gamma$+\oiii$\lambda 4363$. Some improvements are also implemented, including the background residual subtractions, hot pixel removals, optimal 1D extractions, as well as flux calibrations by referring to the standard star observations taken during commissioning. 
Using the improved NIRSpec spectrum, Nakajima et al. confirm the rest-frame UV and optical emission line detection at $z=8.496$ from \targ, including  \oiii$\lambda5007,4959$, \oiii$\lambda4363$, H$\beta$, H$\gamma$, H$\delta$, and \ciii$\lambda\lambda1907,1909$, and evaluate their relative line fluxes 
after applying a slit-loss correction\footnote{
The slit-loss is corrected by convolving and integrating the 1D spectrum with the filter response of F444W to match it with the total magnitude of the NIRCam photometry, which is derived with a $0\farcs3$-diameter circular aperture and the corresponding aperture correction. 
}. 
On the other hand, Nakajima et al. find that the \oii3727 doublet is not clearly detected in every nod and visit, and they place a conservative 3$\sigma$ upper limit, 
which provides a lower limit for the ionization parameter $\log$(U)$>-2.27$ based on \cite{kewley2002}.  
The electron temperature and the gas-phase oxygen abundance are estimated to be $T_{\rm [OIII]}=(2.08\pm0.34)\times10^{4}$~K and 12+$\log$(O/H) = $7.26\pm0.18$ (equal to $0.04\pm0.02\,Z_{\odot}$ by assuming the solar metallicity of 12+$\log$(O/H)=8.69) via the direct $T_{\rm e}$ method \citep[e.g.,][]{pilyugin2012} by assuming a negligible contribution from O$^{+}$/H$^{+}$ ($<0.05$\,dex at $3\sigma$; \citealt{nakajima2023}). 
The lower and upper boundaries of the C/O abundance are also estimated to be $\log$(C/O) = $[-0.52:-0.24]$ \citep{isobe2023b}. 
These physical parameters are derived including the dust correction based on the best-fit $E(B-V)$ value in this paper (Section \ref{sec:full_sed}).

\setlength{\tabcolsep}{1.5pt}
\begin{table}
\begin{center}
\caption{Observed FIR properties of \targ\ with ALMA}
\label{tab:fir_prop}
\vspace{-0.4cm}
\begin{tabular}{ccc}
\hline \hline
Target line & \oiii88$\mu$m & \cii158$\mu$m \\ 
\hline R.A. &  07:23:26.242 & 07:23:26.243  \\
Dec.        & $-73$:26:56.824  &  $-73$:26:56.514 \\
S/N$_{\rm line}$ & 4.0 & 4.5  \\ 
Frequency center [GHz] & $357.305\pm0.034$ & $200.203\pm0.010$ \\
$z_{\rm line}$  &  $8.4963\pm0.0009$    & $8.4931\pm0.0005$  \\ 
Line width [km~s$^{-1}$] & $136\pm67$ & 118 $\pm$ 36 \\
Line intensity [Jy~km~s$^{-1}$] & $0.113\pm0.033^{\sharp}$ & $0.094\pm0.027^{\dagger}$  \\
Line luminosity [$\times10^{8}\,L_{\odot}$] & $3.12\pm0.89^{\sharp}$  & $1.45\pm0.32^{\dagger}$ \\
Continuum [$\mu$Jy]    & $\,\,<41.8$ (0.85~mm) &    $<23.2$ (1.54~mm)  \\
\hline \hline
\end{tabular}
\end{center}
$\sharp$ We add a possible uncertainty by 20\% due to the velocity-integration range for the \oiii\ line (see text).  \\
$\dagger$ Given the spatial and velocity offsets of the \cii\ line, we place a 3$\sigma$ upper limit of $L_{\rm [CII]} < 6.0\times10^{7}\,L_{\odot}$ at the galaxy position from the residual map of {\sc imfit}.  
\end{table}

\begin{figure*}
\includegraphics[trim=0cm 0cm 0cm 0cm, clip, angle=0,width=1\textwidth]{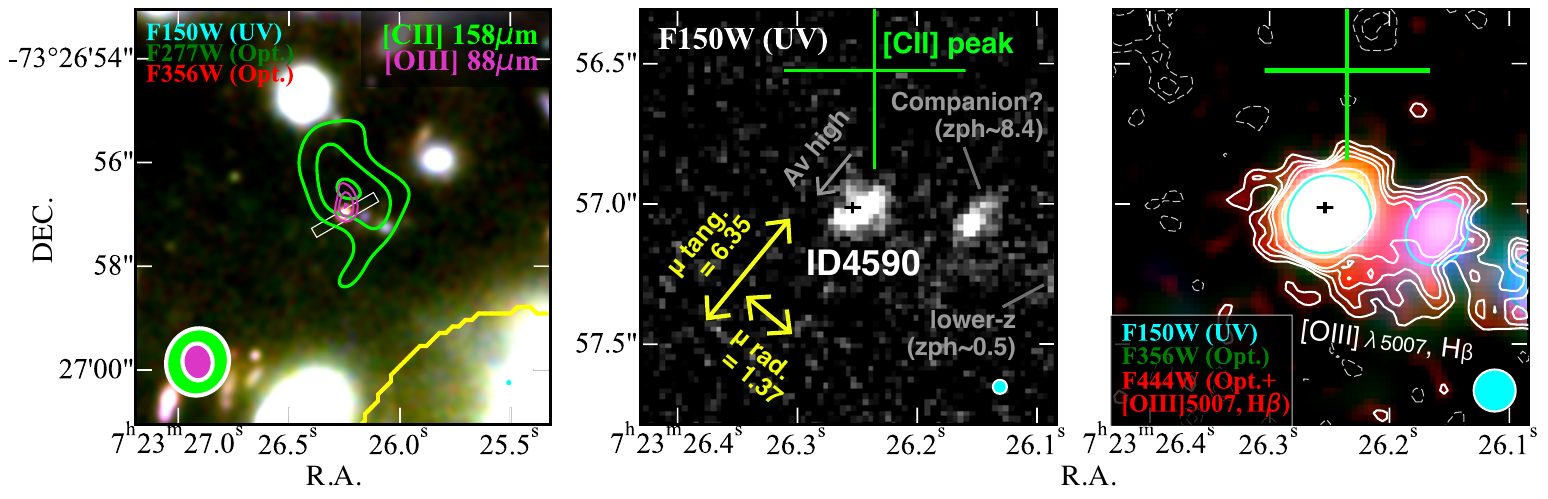}
 \caption{
NIRCam image cutouts around \targ. 
\textbf{\textit{Left:}} 
PSF-matched RGB image ($6''\times6''$) whose color assignment is shown in the label.  
The light green and magenta contours indicate the \cii158$\mu$m and \oiii88$\mu$m line intensities at 2$\sigma$, 3$\sigma$, and 4$\sigma$ levels, respectively. 
The yellow curve denotes the critical curve at $z=8.5$ of SMACS0723 taken from the latest {\sc Glafic} model (Section \ref{sec:model}). 
The light green and magenta ellipses at the bottom left are the ALMA synthesized beams in Band~5 (for \cii) and Band~7 (for \oiii). 
The white rectangle shows the NIRSpec/MSA slit.  
\textbf{\textit{Middle:}}
Zoom-in ($1\farcs5\times1\farcs5$) NIRCam/F150W image. 
The green and black crosses indicate the \cii158$\mu$m and F444W emission peak positions, where the bar scales correspond to their positional accuracy. 
The yellow arrows indicate the radial ($\mu_{\rm rad.}$) and tangential ($\mu_{\rm tang.}$) magnification factors, which combine for a total magnification of 8.69. 
There are two nearby ($\sim0\farcs4$--$0\farcs8$) objects towards the East. 
The SED fitting with \texttt{EAZY} suggests that the closer one could be a companion of \targ\ with $z_{\rm phot}\sim8.4$, while the other is estimated to lie at $z_{\rm phot}\sim0.5$. 
The grey arrow indicates the dust attenuation gradient measured by a pixel-based SED analysis \citep{clara2022b}, implying that the offset of the \cii\ emission is not caused by a dust-obscured star-forming region. 
\textbf{\textit{Right:}} 
Zoom-in ($1\farcs5\times1\farcs5$) PSF-matched RGB image whose color assignment is shown in the label. 
The white contours indicate the flux distribution in the F444W filter at $\pm$2$\sigma$, $\pm$3$\sigma$, $\pm$4$\sigma$, $\pm$5$\sigma$, and $\pm$10$\sigma$ levels. 
For comparison, the cyan contour represents the PSF-matched F150W flux distribution whose intensity relative to the peak is equal to that of the 2$\sigma$-level contour in the F444W filter.  
The F444W filter includes the rest-frame optical continuum and \oiii5007 and H$\beta$ emission line from \targ, showing an extended structure more than other NIRCam filters. 
The NIRCam filters are PSF-matched to the F444W filter in the left and right panels. 
\label{fig:nircam}}
\end{figure*}
\vspace{0.6cm}

\section{Data analysis}
\label{sec:analysis}

\subsection{Mass Model}
\label{sec:model}

Several mass models are publicly available in SM0723, including those developed by using {\sc Lenstool} \citep{jullo2007} and {\sc Glafic} \citep{oguri2010} with the \hst\ data from the RELICS team \citep{coe2019}\footnote{
\url{https://archive.stsci.edu/prepds/relics/}
}. 
More models have been constructed for SM0723 with the \jwst\ ERO data, using {\sc Lenstool} \citep{mahler2022, caminha2022}, the light-traces-mass approach (LTM; \citealt{broadhurst2005, zitrin2009, zitrin2015}) presented in \cite{pascale2022}, and an updated version of {\sc Glafic} \citep{oguri2021b} presented in \cite{harikane2022b}. 

In this paper, we use the latest {\sc Glafic} model. 
The magnification factor for \targ\ is estimated to be $\mu=8.69$, which is consistent with the latest {\sc Lenstool} model prediction ($\mu\sim9$) based on the \jwst\ ERO and the latest MUSE data \citep{caminha2022}. 
We define the systematic uncertainty due to the choice of the mass model by 
\begin{eqnarray}
\Delta\mu\equiv \frac{|\mu_{\rm glafic} - \mu_{\rm other}|}{\mu_{\rm glafic}}, 
\end{eqnarray}
where $\mu_{\rm glafic}$ and $\mu_{\rm other}$ indicate the magnification factor evaluated by the updated version of {\sc Glafic} and other models, respectively. 
Among the mass models constructed with the \jwst\ ERO data, the $\Delta\mu$ value is estimated to be $\sim$15--30\% for \targ. In the following analyses when we correct for the gravitationally lensing effect on \targ, we use $\mu=8.69$ and add a systematic uncertainty from the mass models of 30\%. 
When correcting physical values for lensing, we do not account for differential magnification in the following analysis, as we confirm it to be much less than the above systematic uncertainty across the positions in the galaxy.  
However, when we discuss the intrinsic morphology and size (Section \ref{sec:size} and \ref{sec:morph}), we do consider the difference between the radial ($\mu_{\rm rad.}$) and tangential ($\mu_{\rm tang.}$) 
 magnifications that are estimated to be $\mu_{\rm rad.}=1.37$ and $\mu_{\rm tang.}=6.35$ with position angle (PA) in East of North of $-39.5$~deg.  

\subsection{Continuum \& Line Measurements}
\label{sec:alma_measurement}

In the top panels of Figure \ref{fig:alma}, we show the $6''\times6''$ continuum maps of ALMA Band~7 (left) and Band~5 (right). 
In both bands, we find that the relevant pixels show negative counts. 
We assume that the emission is unresolved in the continuum maps with the current ALMA beam ($\sim0\farcs7$--$1\farcs3$) and place 2$\sigma$ upper limits of 41.8 $\mu$Jy~beam$^{-1}$ and 23.2 $\mu$Jy~beam$^{-1}$ in Band~7 and Band~5, respectively, based on the standard deviation of the maps.  

In the bottom panels of Figure \ref{fig:alma}, we also present the line spectra for the \oiii88$\mu$m (left) and \cii158$\mu$m lines (right). 
In the spectra, we identify positive signals in several consecutive channels at around 357.3~GHz and 200.2~GHz in Band~7 and Band~5, respectively, that are both consistent with the expected frequencies of the \oiii88$\mu$m and \cii158$\mu$m lines based on the source redshift spectroscopically determined by NIRSpec. 
In the top panels, we show the velocity-integrated (moment 0) maps by using these line-detected channels that are highlighted by the yellow shades in the bottom spectra. 
In the moment 0 maps, we find that the S/N in the peak pixel shows 4.0 and 4.1 for the \oiii\ and \cii\ lines, respectively. The latter is spatially resolved, especially in the Northwest to Southeast direction, where the S/N increases to 4.5 with a $2\farcs0$-diameter aperture. 
On the other hand, \oiii\ line is compact and not spatially resolved with the current beam size and data depth. 
Note that the \oiii\ line shows a potential double peak profile, which is typical among rotation-supported systems with inclinations \citep[e.g.,][]{kohandel2019}, while the other possibility is that the noise fluctuation skews the line profile from a single Gaussian. 
We confirm that removing the three channels around the secondary peak ($\simeq$357.45~GHz) from the integration range for the mom~0 map provides the same peak S/N of 4.0. This suggests that the significance level of the \oiii\ line detection is not unchanged regardless of whether it is truly the double peak profile or not.

From a single Gaussian fit in the spectra, we estimate the frequency center and the line redshifts as $z_{\rm [OIII]88}=8.4963 \pm 0.0009$ and $z_{\rm [CII]}=8.4931 \pm 0.0005$ with the line width of FWHM$_{\rm [OIII]88}=137\pm67$~km~s$^{-1}$ and FWHM$_{\rm [CII]}=$  $118\pm36$ km~s$^{-1}$. 
The redshift of \oiii88$\mu$m is in excellent agreement with the source redshift determined by NIRSpec, while the \cii\ line appears blue-shifted by $\sim90$~km~s$^{-1}$ beyond the errors. 
The line widths are consistent with each other within the errors. 

Given a possibility that the chance projection of the noise fluctuation \citep[e.g.,][]{kaasinen2022} causes the \cii\ velocity offset, 
we quantify its probability in our \cii\ line identification by running a blind line search algorithm of {\sc Findclump} implemented in a Python library of {\sc Interferopy} \citep{interferopy} for observational radio--mm interferometry data analysis\footnote{
\url{https://interferopy.readthedocs.io/en/latest/index.html}
}. 
We use a 20-km~$^{-1}$ channel width cube of Band~5 and adopt the spatial tolerance of $1\farcs2$ (= beam size) and the frequency tolerance of 100~MHz ($\sim$150~km~s$^{-1}$) to match the detection in the cube. 
We find that $\sim180$ line features are identified with similar or higher S/N in the entire data cube within a 20$''$-radius circular area and 274 channels,  which is equal to 259793 independent beams in the cube. 
Given that the 200.2~GHz line is identified within the 
spatial and frequency tolerances from the expectations of the target, we estimate the chance projection of the random noise is 0.07\% (= 180 / 259793 $\times$ 1), indicating that the purity of the line detection is $>99.9\%$.  
Therefore, we conclude that \cii\ line identification is unlikely explained by the noise fluctuation, and its velocity and spatial offsets are real. 
The reason for the blue-shifted \cii\ would be differential distributions of the multi-phase gases and their associated kinematics \citep[e.g.,][]{pallottini2019, arata2020, kohandel2020, katz2022, akins2022, valentino2022}; the rest-frame optical emission lines observed with NIRSpec and the \oiii88$\mu$m line originate from the ionized gas, while the \cii158$\mu$m line mostly arises in the photodissociation regions (PDRs). 
The \cii\ peak position shows a spatial offset from the \jwst\ source position by $\sim 0\farcs5$ beyond the uncertainty of its positional accuracy (Section \ref{sec:morph}), which also supports the differential distributions of these multi-phase gases. 

With a 2$''$-diameter aperture in the moment 0 maps, we measure the line intensities and convert them to the \cii\ line luminosity of $L_{\rm [CII]}=(1.45\pm0.32)\times10^{8}\,L_{\odot}$ in the observed frame (i.e., no lens correction). 
Given its spatial and velocity offsets, we also place a $3\sigma$ upper limit at the galaxy position of $< 6.0\times10^{7}\,L_{\odot}$ by taking the line width from \oiii\ line.  
From its compact morphology of \oiii\ 88$\mu$m, we assume that \oiii\ is spatially unresolved and infer the \oiii\ line luminosity of $L_{\rm [OIII]}=(3.12\pm0.76)\times10^{8}\,L_{\odot}$ from the peak pixel count. 
Note that this \oiii\ line flux estimate decreases by $\sim20\%$ if we remove the three channels around the secondary peak from the integration range when generating the moment~0 map. We thus add a systematic uncertainty of 20\% for the \oiii\ line flux estimate in the following analyses.
We summarize these continuum and line properties in Table \ref{tab:fir_prop}. 

\subsection{Sizes}
\label{sec:size}

We also measure the spatial sizes of \cii158$\mu$m and \oiii88$\mu$m lines with ALMA as well as the rest-frame UV and optical continuum with \jwst/NIRCam. In both data, we measure the sizes in the image plane. 

First, for the \cii\ and \oiii\ lines, we perform CASA {\sc imfit} task to apply the 2D elliptical Gaussian fitting. We do not fix any parameters in the fitting. We obtain the best-fit FWHM of $1\farcs90\pm0\farcs82$ and $0\farcs99\pm0\farcs57$ in the major and minor axis, respectively, for the \cii\ line. 
Although we cannot exclude the possibility that these best-fit sizes are affected by the noise fluctuation with the current S/N, we confirm that no significant positive/negative signals remain in the residual map.  
On the other hand, the {\sc imfit} output suggests that \oiii\ is not spatially resolved. We subtract the ALMA synthesized beam profile re-scaled to the \oiii\ peak count from the observed \oiii\ map and confirm that no significant positive/negative signals remain in the residual map. We thus place a $2\sigma$ upper limit of FWHM $<0\farcs31$ based on the limit of the reliable size measurement with interferometric data according to the data sensitivity and the beam size \citep{marti2012}. 
The trend of the larger \cii\ line size than that of the \oiii\ line is consistent with recent ALMA observation results for galaxies at similar redshifts \citep{carniani2020, akins2022,witstok2022}.  
We list the circularized effective radii \footnote{
The effective radius is almost FWHM / 2.0 in Gaussian.  
} of the \cii\ and \oiii\ lines in Table~\ref{tab:property}. 
The observed, best-fit model and residual maps are summarized in the Appendix. 

Second, for the rest-frame UV and optical continuum, we use the NIRCam/F150W and F356W maps, respectively, that are not affected by strong emission line contributions. 
We conduct a 2D S\'ersic profile fitting with {\sc galfit} \citep{peng2010}. 
The pixel of the F356W map is rebinned to a pixel scale of $0\farcs02$, which is the same pixel scale as the F150W map. 
We use NIRCam PSFs for the ERO data of SM0723 publicly available\footnote{
\url{
https://github.com/gbrammer/grizli-psf-library
}
}, which are generated from the \texttt{WebbPSF} model and drizzled to a grid of $0\farcs02$ in the same manner as the mosaic maps in the \texttt{grizli} pipeline.  
Note that these PSFs are generated with the latest version of \texttt{WebbPSF}, including the correction of the optical path difference\footnote{
\url{
https://webbpsf.readthedocs.io/en/latest/jwst_measured_opds.html
}
}, where it mitigates the potential difference from the empirical PSFs \citep[e.g.,][]{ono2022}. 
In the F150W fit for the rest-frame UV continuum, we do not fix any parameters, while we adopt the initial parameter set based on the lensing distortion with the axis ratio of $\mu_{\rm rad.}/\mu_{\rm tang.}=0.22$ and the PA of $-39.5$~deg. We obtain the best-fit effective radius of $r_{\rm e}=3.6\pm0.5$~pixel ($=0.33\pm0.05$~kpc) in the major axis with the axis ratio of $0.57\pm0.09$. 
The best-fit axis ratio exceeds the prediction from the lensing distortion ($=0.22$), indicating that \targ\ is well resolved in the radial magnification axis. 
We also obtain the PA of $-38\pm10$~deg, which is consistent with the lensing distortion ($=-39.5$). This supports the strong magnification factor in \targ\ of $8.69$ ($=\mu_{\rm rad.}\times\mu_{\rm tang.}$) and suggests that the observed rest-frame UV morphology is mostly dominated by the lensing distortion. 
In the F356W fit for the rest-frame optical continuum, we fix the best-fit axis ratio, PA, and Se\'rsic index from the F150W results, given the worse spatial resolution than that of F150W.  
We obtain the best-fit $r_{\rm e}$ of $3.9\pm0.2$~pixel ($=0.36\pm0.02$~kpc) in the major axis. 
Given the best-fit axis ratio of $0.57\pm0.09$, we list the circularized $r_{\rm e}$ values for these stellar continua also in Table~\ref{tab:property}. 
The observed, best-fit model, and residual maps for these NIRCam results are also summarized in the Appendix.

\subsection{Morphology}
\label{sec:morph}

In Figure \ref{fig:nircam}, we compare the spatial distribution of each emission from \targ. 
The left panel shows the \cii158$\mu$m and \oiii88$\mu$m line contours overlaid on the $6''\times6''$ RGB color image with NIRCam filters. 
The middle and right panels display zoom-in ($1\farcs5\times1\farcs5$ ) F150W image (middle) and RGB color image (right) with NIRCam filters of F150W, F356W, and F444W, where the first two filters trace the rest-frame UV and optical continuum, while the last filter includes the \oiii5007 and H$\beta$ lines in addition to the underlying rest-frame optical continuum. 
For the RGB images, the maps are PSF-matched to the F444W filter, and 
the white and cyan contours in the right panel represent the flux distribution in the F444W and F150W filters, respectively. 
The green and black crosses indicate the peak position of the \cii158$\mu$m line and the F444W emission, where the bar sizes of the cross are equal to the uncertainty of the positional accuracy based on its beam size and S/N\footnote{
The positional accuracy of ALMA (pos$_{\rm acc}$) can be approximated by the relationship of $\theta_{\rm beam}$ / SNR / 0.9, 
where SNR is the signal-to-noise ratio of the image target peak (\url{https://help.almascience.org/kb/articles/what-is-the-absolute-astrometric-accuracy-of-alma}).
}.
The yellow arrows in the middle panel denote the radial ($\mu_{\rm rad.}$) and tangential magnifications ($\mu_{\rm tang.}$) to understand the distortion of \targ\ due to the lensing effect. 
To investigate faint tails of the emission, we additionally subtract the local background in all NIRCam filters by evaluating the median pixel count in a black field near from \targ. 

We find that the peak position of the \oiii88$\mu$m is consistent with the \jwst\ source position of \targ, while the \cii158$\mu$m has an offset of $\sim0\farcs5$ beyond the uncertainty of the positional accuracy.
In the source-plane reconstruction of the \jwst\ and \cii\ source positions, we find that the intrinsic offset decreases down to $\sim0\farcs1$, which is equal to $\sim0.5$~kpc. 
Because of the significantly low probability of the chance projection of the noise (Section \ref{sec:alma_measurement}), 
the spatial offset in \cii\ indicates that physical origins of the emission is associated with \targ, while it arises outside the galaxy \citep[e.g.,][]{maiolino2015, carniani2017}. 
We further discuss the physical origins of the \cii\ offset in Section \ref{sec:extended}. 

We also find that the emission in the F444W filter is extended more than the rest-frame UV and optical continuum observed in F150W and F356W filters. 
We interpret this extended emission attributed to strong emission lines of \oiii5007 and H$\beta$ from \targ, implying that a powerful mechanism of forming the extended ionized gas structure is taking place. 
Note that there are two nearby objects towards the East with offsets of $\simeq0\farcs4$ and $\simeq0\farcs8$. 
We run the SED fitting code \texttt{EAZY} \citep{brammer2008} for these two nearby objects with the available \jwst/NIRCam, MIRI, and \hst\ photometry in the public \texttt{grizli} catalog (Section \ref{sec:jwst}) by using the default template set of \texttt{tweak\_fsps\_QSF\_12\_v3}. The results suggest the photometric redshifts of $z_{\rm phot}=8.45^{+0.29}_{-0.28}$ and $z_{\rm phot}=0.37^{+0.11}_{-0.10}$ each from the nearest. 
Therefore, the nearest object might be a companion galaxy associated with \targ.
However, if the presence of nearby objects is the cause of the extended structure, the same structure should also be observed in the other NIRCam filters, which is not the case. 
Besides, the structure is also extended towards South $\sim$ Southeast, where no rest-frame UV continuum is identified. 
Furthermore, the PSF size of the NIRCam/F444W filter is $\sim0\farcs15$, where the emission should be individually resolved if the extended ionized gas is caused by individual further faint satellites. 
We thus conclude that this diffuse, extended structure in the F444W filter is hardly explained either by these two nearby objects or further faint satellites. 
The structure extends to $\sim0\farcs5$ towards the Southeast most. Given the circularized rest-frame optical effective radius of $0\farcs059\pm0\farcs010$ (Section \ref{sec:size}), the structure extends out to $\sim8$ times more than the stellar distribution of the galaxy. 
If we take the differential magnification effects into account, 
the structure is aligned to the radial magnification axis ($\mu_{\rm rad.}=1.37$), indicating that the intrinsic physical distance after the lens correction is $\sim1.7$~kpc. 
For the same direction, the rest-frame optical effective radius after the lens correction is estimated to be $0.11 \pm 0.03$~kpc. 
These results suggest that the ionized gas distribution over the effective radius of the stellar distribution even increases to a factor of $\sim$15.   
The relative ratio of $>8$ is well beyond the diffused ionized gas (DIG) structure observed among local galaxies ($\sim$10\% of the galaxy size; see e.g., \citealt{rossa2003b}).    
We further discuss the physical origins of the extended ionized gas in Section \ref{sec:extended}. 

By comparing the total flux correction factors in F356W and F444W filters (Section \ref{sec:jwst}), we find that the extended component in F444W contributes to the total flux measurement by $\sim8$\%.  
In the following analyses when studying the same emitting regions\footnote{
We correct the 8\% contribution from the extended region in the full SED analysis (Section~\ref{sec:full_sed}). By assuming that \oiii5007 emission also has the same flux contribution from the extended region, we apply the same correction to the \oiii88$\mu$m/\oiii5007 line ratio analysis (Section~\ref{sec:line_ratio}). 
}, we remove this 8\% contribution of the extended ionized gas to the total flux measurement in the F444W filter. 
With the same motivation for analyses that assume the emission originated from the same regions, 
we also use the 3$\sigma$ upper limit for the \cii\ luminosity at the galaxy position {\sc imfit} (Section \ref{sec:alma_measurement}).  

\subsection{FIR SED}
\label{sec:fir_sed}

\begin{figure}
\includegraphics[trim=0cm 0cm 0cm 0cm, clip, angle=0,width=0.5\textwidth]{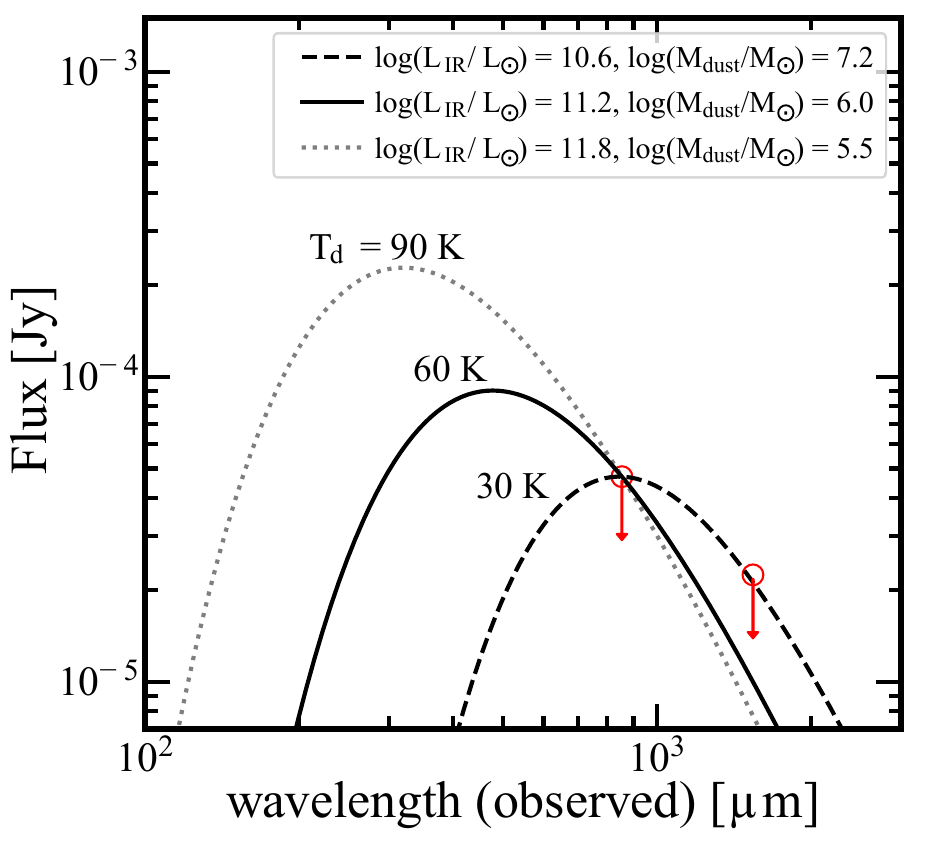}
 \caption{
Constrain of the FIR SED of \targ. 
The red arrows indicate the 2$\sigma$ upper limits obtained in ALMA Band~5 and Band~7. 
The dotted, solid, and dashed curves show the single modified black body (MBB) scaled to the Band~7 upper limit, with the dust temperature of $T_{\rm d}=$ 30~K, 60~K, and 90~K, respectively. 
In each case, the induced IR luminosity $L_{\rm IR}$ by integrating MBB over 8--1000~$\mu$m and the dust mass $M_{\rm dust}$ are shown in the label. We fix the dust spectral index $\beta_{\rm d}=1.8$ \citep[e.g.,][]{chapin2009, planck2011}. 
\label{fig:fir_sed}}
\end{figure}

From the upper limits of the dust continuum both in Band~5 and Band~7, we attempt to constrain the FIR SED of \targ. Recent FIR SED studies for high-$z$ dusty star-forming galaxies at $z\sim1$--4, including \textit{Herschel} and ALMA photometry, suggest the typical dust temperature of $T_{\rm d}\sim30$~K \citep[e.g.,][]{swinbank2014, sun2022}. For UV-selected galaxy populations such as Lyman-break galaxies (LBGs) at $z\sim4$--7, ALMA multiple band observations show a higher dust temperature distribution typically ranging from $\sim$40~K even out to $\sim$80~K \citep[e.g.,][]{faisst2020b, bakx2020, akins2022, witstok2022}, where several analytical models have been developed and well reproduced the observation results, including the potential redshift evolution of $T_{\rm d}$ \citep[e.g.,][]{inoue2020, fudamoto2022a, sommovigo2022}. 

Because of the limited constraints due to lack of the detection in both ALMA Bands, we assume the single modified black body (MBB) for the FIR SED of \targ. We adopt a fiducial $T_{\rm d}$ estimate of 60~K based on the extrapolation of the best-fit redshift evolution model of $T_{\rm d}$ following the decrease of the gas depletion time scale ($t_{\rm depl}$) derived in \cite{sommovigo2022}, while we include the uncertainty of $T_{\rm d}$ by 30~K given the $T_{\rm d}$ distribution so far observed in high-$z$ LBGs with ALMA. We fix dust spectral index $\beta_{\rm d}$ at a typical value of 1.8 \citep{chapin2009, planck2011} and take the CMB temperature effect \citep[e.g.,][]{dacunha2013} into account in the MBB model.

In Figure \ref{fig:fir_sed}, we show three MBB models with $T_{\rm d}=30$~K, 60~K, and 90~K normalized to the upper limit of Band~7. 
The inset labels show the derived IR luminosity $L_{\rm IR}$, integrated over 8--1000~$\mu$m, and the dust mass $M_{\rm dust}$, with the dust opacity coefficient of $\kappa_{\nu}=5.1(\nu/\nu_{250\mu{\rm m}})^{\beta_{\rm d}}$. 
In the observed frame (i.e. no lens correction), we obtain the upper limits of $\log(L_{\rm IR}/L_{\odot})<11.2$ and $\log(M_{\rm dust}/M_{\odot})<6.0$ with the fiducial $T_{\rm d}$ value of 60~K. 
This indicates that \targ\ has the observed-frame dust-obscured SFR of SFR$_{\rm IR}<16\,M_{\odot}$~yr$^{-1}$. 
The FIR SED properties are summarized in Table~\ref{tab:property}.

\begin{figure*}
\includegraphics[trim=0cm 0cm 0cm 0cm, clip, angle=0,width=1\textwidth]{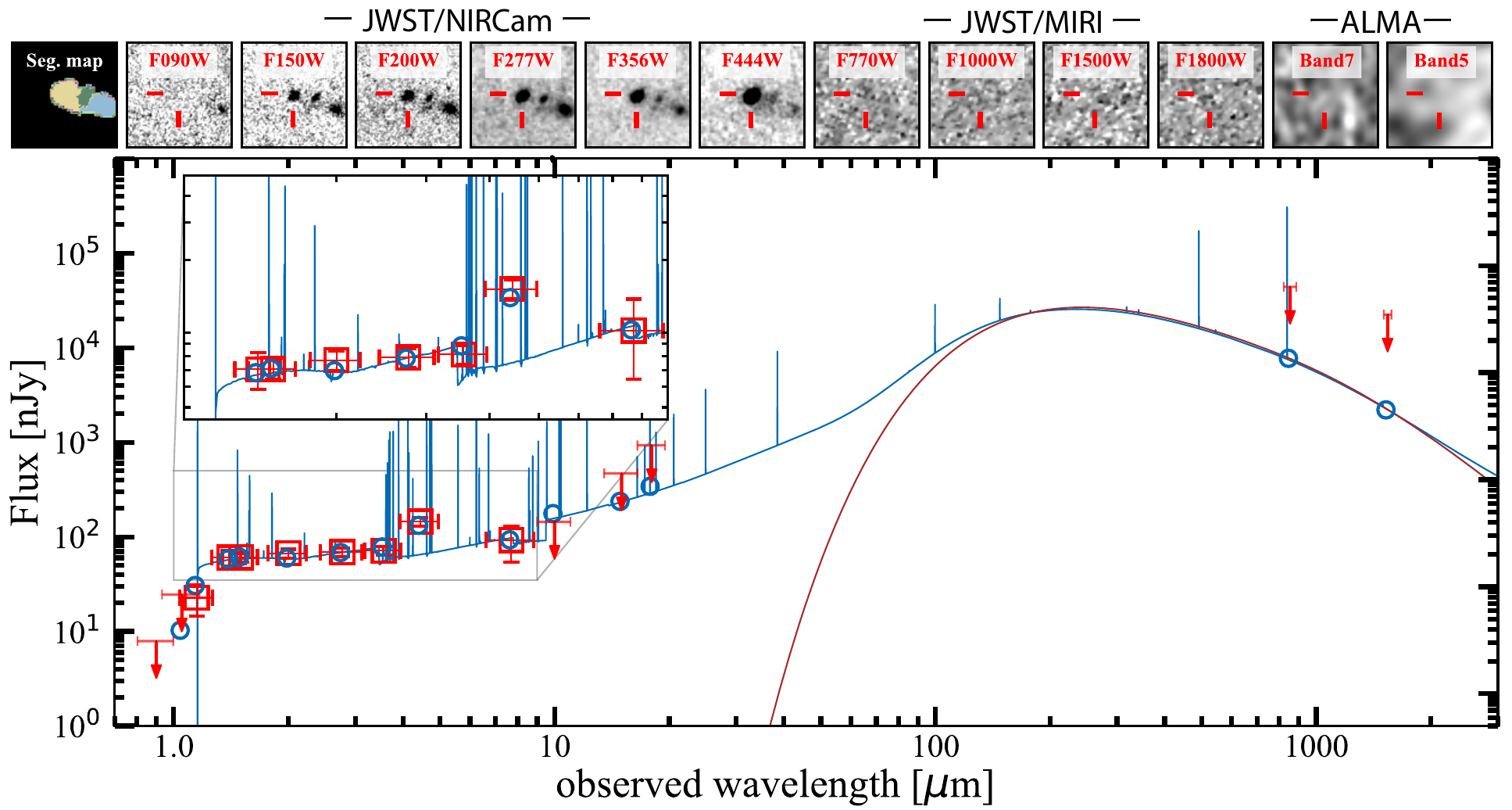}
 \caption{
Full SED shape of \targ\ from optical to mm wavelengths in the observed frame. 
\textbf{\textit{Top}}: 
$2''\times2''$ image cutouts of \jwst/NIRCam, \jwst/MIRI, and ALMA from left to right. 
The red bars indicate the source postion of \targ, and the corresponding filter/Band in each panel is shown in the label. 
The most-left panel shows the segmentation map from the {\sc grizli} pipeline.
\textbf{\textit{Bottom}}: 
Blue curve indicates the best-fit SED obtained with \texttt{CIGALE}. 
The red squares are the observed photometry of \targ, and the blue circles present the predicted photometry from the best-fit SED.
The inset panel zooms in the SED at 1--9~$\mu$m, clearly showing the flux boosting in the F444W filter by strong \oiii$\lambda$5007 and H$\beta$ emission lines, instead of the stellar continuum, where the MIRI photometry helps to disentangle their contributions to the F444W filter. 
The inset panels also presents the red UV continuum slope ($\beta_{\rm UV}=-1.7\pm0.07$) of \targ, while the dust continuum is not detected likely because of the high dust temperature (Section \ref{sec:irx}). 
The brown curve highlights the FIR SED using the \cite{casey2012} model in \texttt{CIGALE}.  
\label{fig:full_sed}}
\end{figure*}

\subsection{Full SED analysis and physical properties}
\label{sec:full_sed}

\setlength{\tabcolsep}{20pt}
\begin{table}
\begin{center}
\caption{Physical properties of \targ\ in the image plane}
\vspace{-0.4cm}
\label{tab:property}
\begin{tabular}{lc}
\hline 
\hline
Name      &    \targ\  \\ \hline
R.A.      & 07:23:26.255       \\ 
Dec.      & $-73$:26:57.041     \\ 
$z_{\rm spec}$   &  8.496 \\
$\mu^{\dagger}$            & 8.69    \\ 
SFR $\times\mu$ [$M_{\odot}$~yr$^{-1}$] &   26$^{+9}_{-4}$     \\
$M_{\rm star}\times\mu$ [$10^{8}\, M_{\odot}$] & 5.3$^{+5.6}_{-2.6}$  \\
$L_{\rm UV}\times\mu$     [10$^{10}$ $L_{\odot}$]  & $2.8\pm0.3$  \\ 
$\beta_{\rm UV}$    & $-1.70\pm0.07$ \\ 
$E(B-V)$            & $0.16\pm0.03$ \\ 
$L_{\rm IR}^{\dagger\dagger}\times\mu$   [$10^{11}\,L_{\odot}$]  &  $<1.6$ \\
$M_{\rm dust}^{\dagger\dagger}\times\mu$  [$10^{6}\,M_{\odot}$]  &  $<1.0$  \\
$M_{\rm gas}\times\mu$  [$10^{9}\,M_{\odot}$]  &  $7\pm3^{a}$ \\
$r_{\rm e, UV}^{\natural}\times \sqrt{\mu}$ [arcsec]       & $0.056\pm0.012$ \\
$r_{\rm e, opt}^{\natural}\times \sqrt{\mu}$ [arcsec]      & $0.059\pm0.010$ \\
$r_{\rm e, [CII]158}^{\natural}\times \sqrt{\mu}$ [arcsec]    & $0.69 \pm0.42$ \\
$r_{\rm e, [OIII]88}^{\natural}\times \sqrt{\mu}$ [arcsec] & $< 0.16$ \\
$n_{\rm e}$ [cm$^{-3}$] & 220$^{+230}_{-130}$ \\
$T_{\rm e}$ [10$^{4}$ K]         & $2.08\pm0.34^{b}$  \\
12 + $\log$(O/H) &  $7.26 \pm 0.18^{b}$\\
$\log(U)$                   & $>-2.27^{b}$ \\
$\log$(C/O)                 & $[-0.52:-0.24]^{c}$ \\
\hline
\end{tabular}
\end{center}
$\dagger$ From the mass model with {\sc Glafic}  (\citealt{oguri2010, oguri2021b}) using the \jwst\ ERO data presented in \cite{harikane2022b}. For the lens corrected values, we add a systematic uncertainty of 30\% throughout the paper (Section \ref{sec:model}).  \\ 
${\dagger\dagger}$ Based on the single MBB with $T_{\rm d}=60$~K and $\beta_{\rm d}=1.8$ without the energy balance, where a $T_{\rm d}$ difference by $\pm30$~K changes the estimates by $\sim\pm$0.5--1.0~dex (see also Figure \ref{fig:fir_sed}). 
\\
$\natural$ Circularized effective radius. The best-fit axis ratio is $0.57\pm0.09$.  \\
$a$ \cite{heintz2022c} \\
$b$ \cite{nakajima2023} \\
$c$ \cite{isobe2023b} 
\end{table}

Figure \ref{fig:full_sed} shows the \jwst\ and ALMA image cutouts, including the segmentation map (top), and the optical--mm photometry in the observed frame (i.e. no lens correction) measured with \hst, \jwst, and ALMA for \targ\ (bottom). 
A significant flux enhancement is observed in the F444W filter (rest-frame $\sim$4000--5000${\rm \AA}$). 
Some remarkably massive early galaxy candidates have been reported at $z\sim7$--11 whose SED shape shows a secondary peak, likely because of the strong Balmer break in the NIRCam LW filters \citep{labbe2022}. 
However, the secondary peak might be explained by contributions from strong emission lines of \oiii$\lambda$5007+H$\beta$ \citep{endsley2022b}, where the lack of the longer wavelength data challenges drawing a definitive conclusion.
In contrast, the presence of the MIRI photometry in \targ\ shows the flux enhancement only occurs in the F444W filter, 
which helps conclude that the flux enhancement in the F444W filter is caused by strong \oiii+H$\beta$ emission lines, rather than a large stellar mass.
Such strong contributions of the \oiii+H$\beta$ lines also agree with the extended morphology observed in the F444W filter, which is interpreted as the presence of the extended ionized gas emission (Section \ref{sec:morph}). 

To perform a panchromatic characterization of \targ, 
we perform SED fitting to the optical--mm photometry using {\sc cigale} \citep{burgarella2005, noll2009, boquien2019}. 
While we examine the FIR SED in Section \ref{sec:fir_sed}, the SED modelling with {\sc cigale} allows us to take the energy balance between the dust absorption and re-emission into account, which is complementary with the independent FIR SED analysis.  
The fitting was performed similarly as in \cite{fujimoto2022b}, and we summarize the details of the fitting and parameter ranges used in the fitting in the Appendix.   

In the bottom panel of Figure \ref{fig:full_sed}, the blue curve shows the best-fit SED, where the brown curve highlights the re-emission of the dust in the rest-frame FIR wavelength based on the \cite{casey2012} model. 
The inset panel shows the zoom-in spectrum of the best-fit SED at $\sim$1--9~$\mu$m. 
Our best-fit SED reproduces the observed photometry, including the flux enhancement in the F444W filter with the reduced $\chi^{2}$ value of 0.99. 
We obtain the best-fit values of SFR = 26$^{+9}_{-4}\,M_{\odot}$~yr$^{-1}$, $M_{\rm star}=5.3^{+5.6}_{-2.6}\times10^{8}\, M_{\odot}$, UV continuum slope $\beta_{\rm UV}=-1.70\pm0.07$, and $E(B-V)=0.16\pm0.03$ in the observed-frame (i.e. no lens correction), respectively, that are generally consistent with previous NIRcam and/or NIRspec based measurements \citep[e.g.,][]{carnall2022, schaerer2022, tacchella2022}, as well as an independent NIRCam and NIRISS based measurements in the separated paper of \cite{heintz2022c}. 
The $\log(L_{\rm IR}/L_{\odot})$ value is estimated to be $11.2$, which is consistent with the upper limit obtained from the independent FIR SED analysis (Section \ref{sec:fir_sed}). 
This indicates that the current non-detection of the dust continuum with ALMA does not violate the energy balance with the SMC dust attenuation curve, while the $\beta_{\rm UV}$ and $E(B-V)$ values suggest that \targ\ is a 
certainly dust-attenuated system, in contrast to very blue galaxies observed in recent \jwst\ observations at similar redshifts \citep[e.g.,][]{nanayakkara2022, furtak2022, cullen2022, topping2022b, finkelstein2023, atek2022, fujimoto2022b, robertson2022}.  
Note that we confirm that the SED outputs are unchanged beyond 5--10\% with and without the ALMA upper limits in the above SED fitting with {\sc cigale}.
We further examine the dust and the obscured properties of \targ\ in Section \ref{sec:irx} and discuss the potential underlying physical mechanisms in Section \ref{sec:extended}. 

Because several Balmer emission lines are detected in \targ\ with NIRSpec, we compare our best-fit $E(B-V)$ with the measurement from the Balmer decrement. The line flux measurements from the latest NIRSpec reduction and calibration (Section \ref{sec:jwst}) 
yield $E(B-V)=0.07^{+0.10}_{-0.07}$ and $0.24^{+0.07}_{-0.07}$ from H$\gamma$/H$\beta$ and H$\delta$/H$\beta$, respectively, by assuming the SMC dust attenuation law. 
Our SED-based estimate falls between those estimates from the Balmer decrement approach, suggesting the general consistency between the photometry- and spectroscopic-based approaches. However, caution remains in the different $E(B-V)$ suggested between H$\gamma$/H$\beta$ and H$\delta$/H$\beta$, where the difference can change the H$\beta$-based SFR estimate \citep[e.g.,][]{kennicutt2012} by a factor of $\sim$3 after the dust correction. We confirm that similarly different $E(B-V)$ values are obtained by assuming other dust attenuation laws (e.g., LMC, \citealt{calzetti2000}), and thus the difference is unlikely caused by an improper choice of the dust attenuation law. 
We speculate that it is caused by the difficulty of the optimal wavelength-dependent slit-loss correction for each emission, taking their potential differential distributions. We thus use the SED-based physical properties in the following analysis.

We note that \cite{clara2022b} discuss the potential underestimate of $M_{\rm star}$ by $\sim$0.5--1~dex in the spatially-integrated SED analysis, compared to the sum of the spatially-resolved SED analysis, especially in strong optical emission line systems 
\footnote{
This is because the young bursty stellar populations,  causing the strong optical emission lines, dominate the emission in the spatially-integrated photometry, 
hiding the possible presence of underlying older stellar populations, 
and yield extremely young ages ($<10$~Myr) and therefore lower masses in the SED fit.
}. 
However, we confirm that our $M_{\rm star}$ estimate is consistent with the results from the spatially-resolved SED analysis in \cite{clara2022b}, owing to the additional constraints from the MIRI bands \citep{bisigello2019, papovich2023}.

\begin{figure*}
\includegraphics[trim=0cm 0cm 0cm 0cm, clip, angle=0,width=1\textwidth]{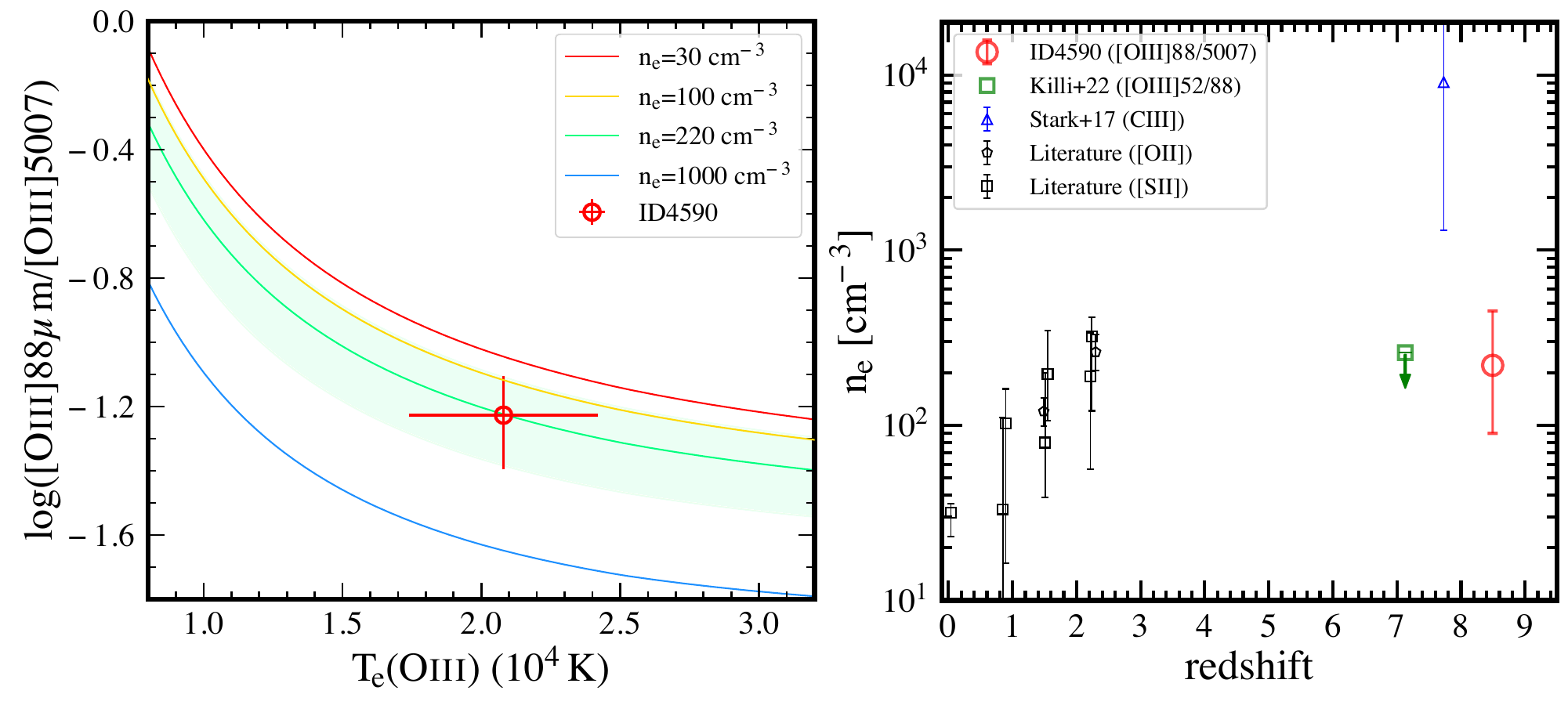}
 \caption{
Electron density $n_{\rm e}$ measurement for \targ\ via the \oiii88$\mu$m/$\lambda$5007 line ratio enabled by \jwst\ and ALMA. 
In both panels, the red circles denote \targ. 
\textbf{\textit{Left:}}
Line ratio as a function of electron temperature $T_{\rm e}$ with different $n_{\rm e}$ assumptions, calculated with the nebular emission code \texttt{PyNeb}. 
The green shade presents the 1$\sigma$ range of the $n_{\rm e}$ estimate for \targ. 
\textbf{\textit{Right:}}
Redshift evolution of $n_{\rm e}$. 
The green and blue symbols are the measurements for individual galaxies at the EoR \citep{killi2022, stark2017}, 
and the black symbols are statistical measurements at $z\sim$0--3 taken from literature \citep{steidel2014, sanders2016, stott2016, kaasinen2017, davies2021}. 
\label{fig:ratio}}
\end{figure*}

\section{Results \& Discussions}
\label{sec:result}

\subsection{Electron density at $z=8.5$ via \oiii88$\mu$m/\oiii$\lambda$5007}
\label{sec:line_ratio}

Owing to the same species and ionized state with different critical densities, the line ratio of \oiii88$\mu$m/\oiii$\lambda$5007 is regulated by the electron density $n_{\rm e}$ and temperature $T_{\rm e}$. 
In the left panel of Figure \ref{fig:ratio}, we show the line ratio of \oiii88$\mu$m/\oiii$\lambda$5007 as a function of $T_{\rm e}$ with different assumptions for the electron density $n_{\rm e}$, drawn by using the nebular emission code \texttt{PyNeb}\footnote{
\url{
http://research.iac.es/proyecto/PyNeb//}
}.
The relations show monotonic decreasing functions towards the increase of $T_{\rm e}$ or $n_{\rm e}$. 
This indicates that we can evaluate $n_{\rm e}$ with a secure $T_{\rm e}$ measurement (or 
vice versa -- evaluate $T_{\rm e}$ with a secure $n_{\rm e}$ measurement). 

The NIRSpec observations successfully detect the multiple nebular emission lines from \targ\ in rest-frame UV to optical wavelength, including \oiii$\lambda$5007 and \oiii$\lambda$4363, which provides us with the robust measure of $T_{\rm e}$ (e.g., \citealt{schaerer2022, curti2022, trump2022}; \citealt{heintz2022c}; \citealt{nakajima2023}). 
In conjunction with the secure $T_{\rm e}$ measurement and our ALMA measurements (Section \ref{sec:alma_measurement}),  
we derive the \oiii88$\mu$m/$\lambda$5007 line ratio for \targ, and the red circle in the left panel of Figure \ref{fig:ratio} shows the relation.  
Based on the 1$\sigma$ uncertainties, we evaluate $n_{\rm e}=220^{+230}_{-130}$~cm$^{-3}$. 

The $n_{\rm e}$ value has been typically measured by using density-sensitive line ratios such as \sii6716/6731, \oii3729/3726, and \ciii1907/1909 
\citep[e.g.,][]{kewley2019}. 
Previous spectroscopic surveys have 
found the presence of the redshift evolution of $n_{\rm e}$: the typical $n_{\rm e}$ in local galaxies has increased from $n_{\rm e}\simeq30$~cm$^{-3}$ at $z\sim0$ \citep[e.g.,][]{herrera-camus2016} to $n_{\rm e}\simeq$100--200~cm$^{-3}$ at $z\sim1.5$ \citep[e.g.,][]{kaasinen2017, kashino2017}, and to $n_{\rm e}\simeq$ 200--300~cm$^{-3}$ at $z\sim2$--3 \citep{steidel2014, sanders2016, davies2021}. 

Because of the required high spectral resolution and subsequently high sensitivity to resolve those rest-frame UV and optical doublet lines, the results have been generally limited at $z\lesssim3$, while recent rest-frame FIR observations have been exploring the $n_{\rm e}$ measurement even out to the EoR. 
By using a FIR fine-structure line ratio of \oiii52$\mu$m and \oiii88$\mu$m detected with ALMA, \cite{killi2022} estimate $n_{\rm e}<260$~cm$^{-3}$ in a dusty lensed star-forming galaxy at $z=7.13$, A1689-zD1 \citep{watson2015}. 
Our measurement of a high electron density provides a new determination of $n_{\rm e}$ at EoR and is consistent with the results in A1689-zD1. 
For several $z\sim6$--9 galaxies with the \oiii88$\mu$m and \cii158$\mu$m line measurements, \cite{vallini2021} show that their gas density $n_{\rm gas}$ generally fall within $\simeq100$--1000~cm$^{-3}$ by advancing analytical models for these FIR emission lines \citep{ferrara2019, vallini2020}.  
Although systematic uncertainties remain in this approach due to the C/O abundance and the different emitting regions of the \oiii\ and \cii\ lines that suggest $n_{\rm e}$ and $n_{\rm gas}$ are not identical, these $n_{\rm gas}$ measurements at similar redshift also in line with our $n_{\rm e}$ measurement for \targ. 
Our measurement is also broadly consistent with recent NIRSpec measurements for $z\sim4-9$ galaxies using the \oii\ doublet \citep{isobe2023}. 
In contrast, \cite{stark2017} use the \ciii\ doublet and estimate $n_{\rm e}=9100^{+12200}_{-7800}$~cm$^{-3}$ in EGS-zs8-1, a UV-luminous star-forming galaxy at $z=7.73$, which is much higher than those of our and recent measurements with the \oiii88$\mu$m line.

In the right panel of Figure \ref{fig:ratio}, we summarize the $n_{\rm e}$ measurement as a function of redshift. 
The $n_{\rm e}$ value of \targ\ is similar to $z\sim2$--3 star-forming galaxies. This might suggest that the physical mechanisms responsible for driving the high $n_{\rm e}$ values observed at $z\sim$2--3 initially took place in the EoR, and there was a little redshift evolution between $z=8.5$ and $z\sim$2--3, albeit the diversity observed with EGS-zs8-1.
However, this redshift trend strongly depends on the galaxy types selected, and similarly high $n_{\rm e}$ measurements ($\simeq400$~cm$^{-3}$) are also obtained in compact star-forming galaxies at $z\sim0.3$--0.4 \citep{guseva2020}.
Moreover, given the similar high ionization potentials between \ciii\ and \oiii88$\mu$m emission lines, but much lower critical density of \oiii88$\mu$m ($\sim500$~cm$^{-3}$), the different $n_{\rm e}$ results between the \ciii\ and \oiii88$\mu$m measurements might indicate these emission lines generally arise from different regions with different $n_{\rm e}$ values, where we are witnessing the differential $n_{\rm e}$ distributions inside {H {\sc ii}} regions.
The upcoming \jwst/NIRSpec observations in the high-resolution spectrograph mode ($R\sim2700$) will sufficiently resolve the rest-frame UV and optical doublet lines, statistically evaluate $n_{\rm e}$ based on a mass-complete sample, verify the presence of its redshift evolution and/or differential $n_{\rm e}$ distributions traced by different emission lines, and determine what key mechanisms regulate $n_{\rm e}$ out to the EoR.

\begin{figure*}
\includegraphics[trim=0cm 0cm 0cm 0cm, clip, angle=0,width=1\textwidth]{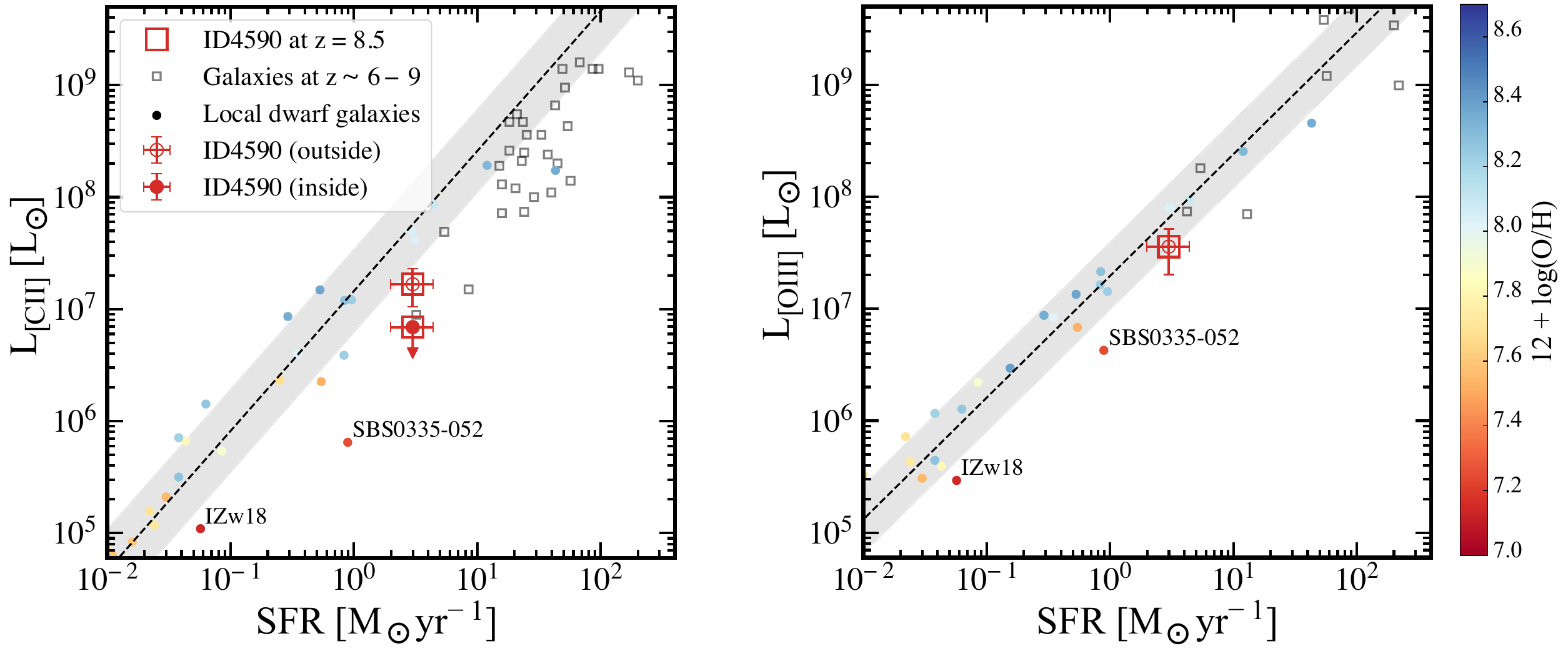}
 \caption{
Relations between SFR and $L_{\rm [CII]}$ (left) and $L_{\rm [OIII]}$ (right). In both panels, the color of the circles corresponds to the gas-phase metallicity $Z_{\rm gas}$ denoted in the color bar. 
The open red squares remark the measurements of \targ, where the filled and open circles indicate the measurements inside and outside the galaxy. 
The red arrows present the 3$\sigma$ upper limit of \cii\ at the \jwst\ source position given its spatial and velocity offsets (Section \ref{sec:alma_measurement}). 
We assume that the \cii-detected gas outside the galaxy is illuminated by ionizing photons escaping from the galaxy and thus use the same SFR value for both results of inside and outside the galaxy.  
Other color circles represent the local dwarf galaxies with the $Z_{\rm gas}$ measurements \citep{delooze2014, cormier2015}, where the best-fit relation and its 1$\sigma$ dispersion is shown with the dashed line and the grey-shaded area, respectively. 
The open black squares show the results for $z\sim6$--9 galaxies compiled in \cite{harikane2020}. 
\label{fig:sfr_line}}
\end{figure*}

\begin{figure}[h!]
\includegraphics[trim=0cm 0cm 0cm 0cm, clip, angle=0,width=0.5\textwidth]{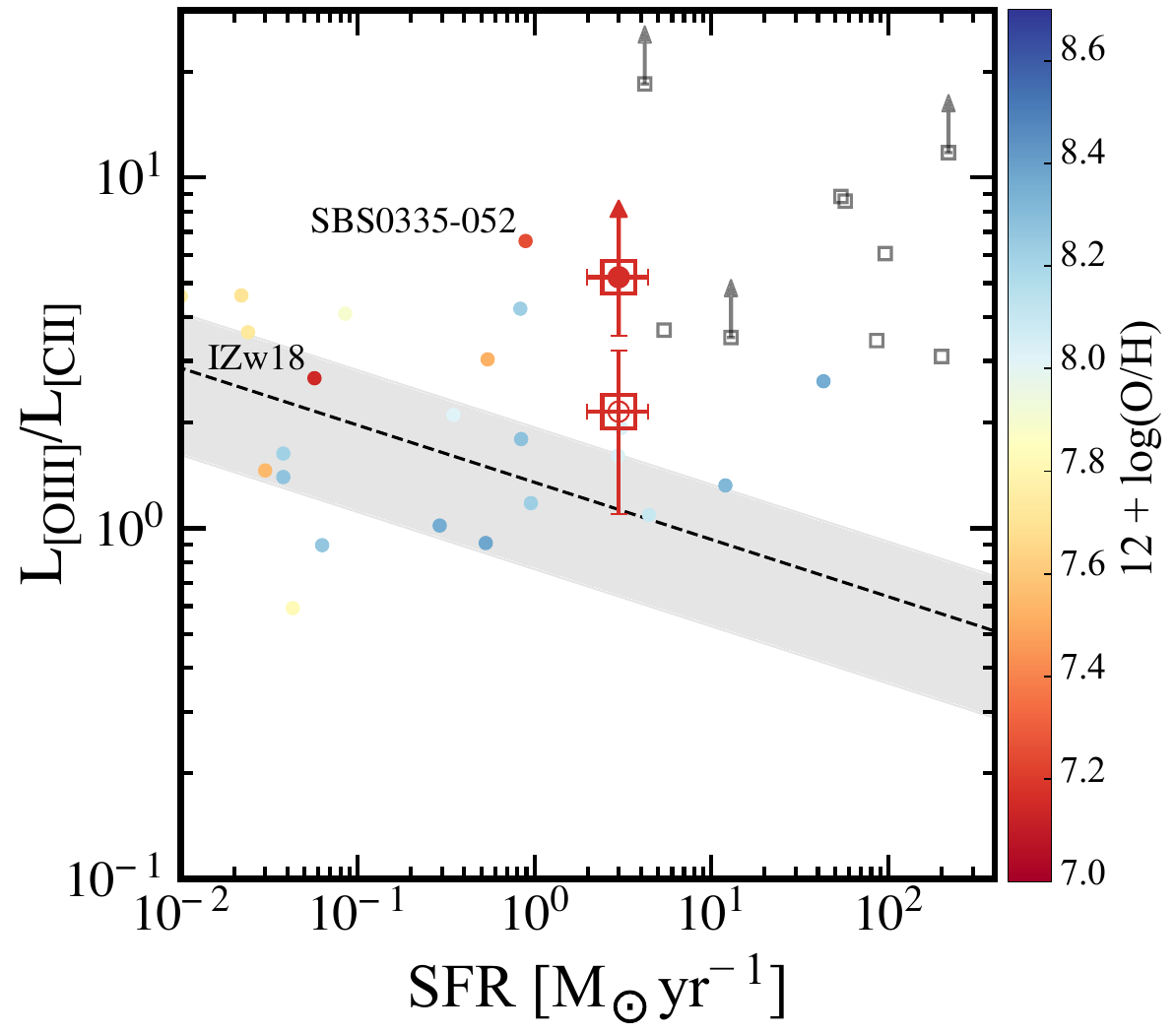}
 \caption{
 Same as Figure \ref{fig:sfr_line}, but for $L_{\rm [OIII]}/L_{\rm [CII]}$. 
 \targ\ shows a high $L_{\rm [OIII]}/L_{\rm [CII]}$ ratio of $>4$ at the galaxy position similar to those of known $z\sim6-9$ galaxies.  
\label{fig:sfr_line2}}
\end{figure}

\subsection{SFR--$L_{\rm [CII]}$, $L_{\rm [OIII]}$, and $L_{\rm [OIII}/L_{\rm [CII]}$ Relations}
\label{sec:cii-oiii}

In Figure \ref{fig:sfr_line} and \ref{fig:sfr_line2}, we present $L_{\rm [CII]}$, $L_{\rm [OIII]}$, and $L_{\rm [OIII]}/L_{\rm [CII]}$ as a function of SFR for \targ. 
We show the results after the lens correction, where the systematic uncertainty of 30\% for the magnification factor (Section \ref{sec:model}) is propagated in the error bars. 
In SFR--$L_{\rm [CII]}$ (SFR--$L_{\rm [OIII]}/L_{\rm [CII]}$) relation, we also show the $3\sigma$ upper limit (lower limit) for $L_{\rm [CII]}$ ($L_{\rm [OIII]}/L_{\rm [CII]}$) at the \jwst\ source position, given its spatial offset (Section \ref{sec:morph}), 
and separate the results inside and outside the galaxy with filled and open circles, respectively. 
For comparison, we also present compilations of recent ALMA results for $z\sim6$--9 galaxies in literature \citep[e.g.,][]{harikane2020, fujimoto2021} with black squares, and local dwarf galaxy results with the gas-phase metallicity ($Z_{\rm gas}$) measurements \citep{delooze2014, cormier2015} with color circles. 
The color scale is equal to 12+$\log$(O/H) denoted in the color bar, except for the $z\sim6$--9 galaxy results whose $Z_{\rm gas}$ values have not been constrained. 
The average relation for the local dwarf galaxies is shown in the dashed black line with the 1$\sigma$ range in the grey shade. 
In all relations, \targ\ are generally consistent with the results estimated in other $z\sim6$--9 galaxies. In addition, \targ\ explores the faint-end of the $z\sim$6--9 galaxy results, owing to the aid of the gravitational lensing effect. 
Therefore, \targ\ is a faint, thus abundant, and representative early galaxy with physical properties similar to other galaxies so far observed with ALMA at similar redshifts. 
This indicates that \targ\ is a unique laboratory to study what regulates the FIR major coolant lines of \cii158$\mu$m and \oiii88$\mu$m in EoR galaxies.   

In the SFR--$L_{\rm [CII]}$ relation, 
we find that \targ\ falls below the typical relation of the local dwarf galaxies in both results obtained inside and outside the galaxy. 
Among the local dwarf galaxies, there are two galaxies of IZw18 and SBS0335-052 whose $Z_{\rm gas}$ measurements are similarly low as \targ\ (12+$\log$(O/H) $\simeq$ 7.1--7.3). 
We confirm that the $\log(L_{\rm [CII]}$/SFR) ratio inside the galaxy shows $<6.3$, which is consistent with these two very metal-poor local galaxies ($\sim5.7$). 
This might indicate that the low \cii\ line emissivity of \targ\ is explained by the low $Z_{\rm gas}$ ISM condition \citep[e.g.,][]{vallini2015}, while there are also other key factors which reduce the \cii\ emissivity such as high ionization parameter or low gas density (see Section \ref{sec:intro}). 
In fact, \targ\ shows strong \oiii$\lambda$5007+H$\beta$ emission (Section \ref{sec:full_sed}), which generally represents recent young bursty stellar populations \citep[e.g.,][]{topping2022, witstok2022}, where the high ionization parameter, as a result, might be a more critical driver. 
The \cii\ emission outside the galaxy shows $\log(L_{\rm [CII]}$/SFR)$\sim6.7$, being higher than these very metal-poor local galaxies but still lower than the typical relation. 
This might be explained by a huge amount of gas around \targ\ which efficiently uses all the photons to boost the \cii\ emission eventually. 
With an analytical model, \cite{ferrara2019} predict that the surface density of \cii\ luminosity becomes almost constant around $\Sigma_{\rm [CII]}\approx10^{6}-10^{7}\,L_{\odot}$~yr$^{-1}$~kpc$^{-1}$
at a high SFR surface density regime of $\Sigma_{\rm SFR}\gtrsim10\,M_{\odot}$~yr$^{-1}$~kpc$^{-2}$ with a linear scale dependence on the gas density, regardless of $Z_{\rm gas}$ in 0.1--1.0 $Z_{\odot}$ (see eq.~42 in \citealt{ferrara2019}). 
Based on our size and full SED analyses in Section \ref{sec:analysis}, \targ\ indeed shows $\Sigma_{\rm [CII]}\simeq2\times10^{6}\,L_{\odot}$~yr$^{-1}$~kpc$^{-2}$ with $\Sigma_{\rm SFR}\simeq100\,M_{\odot}$~yr$^{-1}$~kpc$^{-2}$, which agrees with the prediction from the analytical model. For more discussions related to the rich gas aspect around \targ, we refer the reader to a separate paper by \cite{heintz2022c}.

In the SFR--$L_{\rm [OIII]}$ relation, 
we find that \targ\ is consistent with the local relation within the errors but likely falls slightly below it.  
This would also be explained by the low \oiii\ line emissivity with low $Z_{\rm gas}$ values \citep[e.g.,][]{popping2022}, which is also shown in the monotonic decreasing function in the \oiii88/\oiii$\lambda$5007--$T_{\rm e}$ relation, regardless of $n_{\rm e}$ (Figure \ref{fig:ratio}). 
In fact, the slightly low $\log(L_{\rm [OIII]}$/SFR) ratio of \targ\ is consistent with those of the two very metal-poor local galaxies of IZw18 and SBS0335-052 ($\sim6.5$) within the errors.
We also confirm that the SFR-$L_{\rm [OIII]}$ relation of \targ\ is consistent with the {\sc serra} zoom-in simulation results \citep[e.g.,][]{pallottini2022,kohandel2022} for galaxies whose $\log(U)$ values are similar to that of \targ\ ($\log(U)>-2.27$; Section \ref{sec:jwst}).

In the SFR--$L_{\rm [OIII]}/L_{\rm [CII]}$ relation, 
we find that \targ\ shows higher $L_{\rm [OIII]}/L_{\rm [CII]}$ ratios both inside and outside the galaxies than the local relation. 
In particular, the lower limit of $>4$ obtained inside the galaxy is similarly high to other $z\sim6$--9 galaxies so far observed \citep[e.g.,][]{harikane2020, witstok2022}. 
We discuss the physical origins of the high \oiii/\cii line ratio in Section \ref{sec:oiii-cii_ratio}.


\subsection{IRX-$\beta$ relation}
\label{sec:irx}

\begin{figure}
\includegraphics[trim=0cm 0cm 0cm 0cm, clip, angle=0,width=0.47\textwidth]{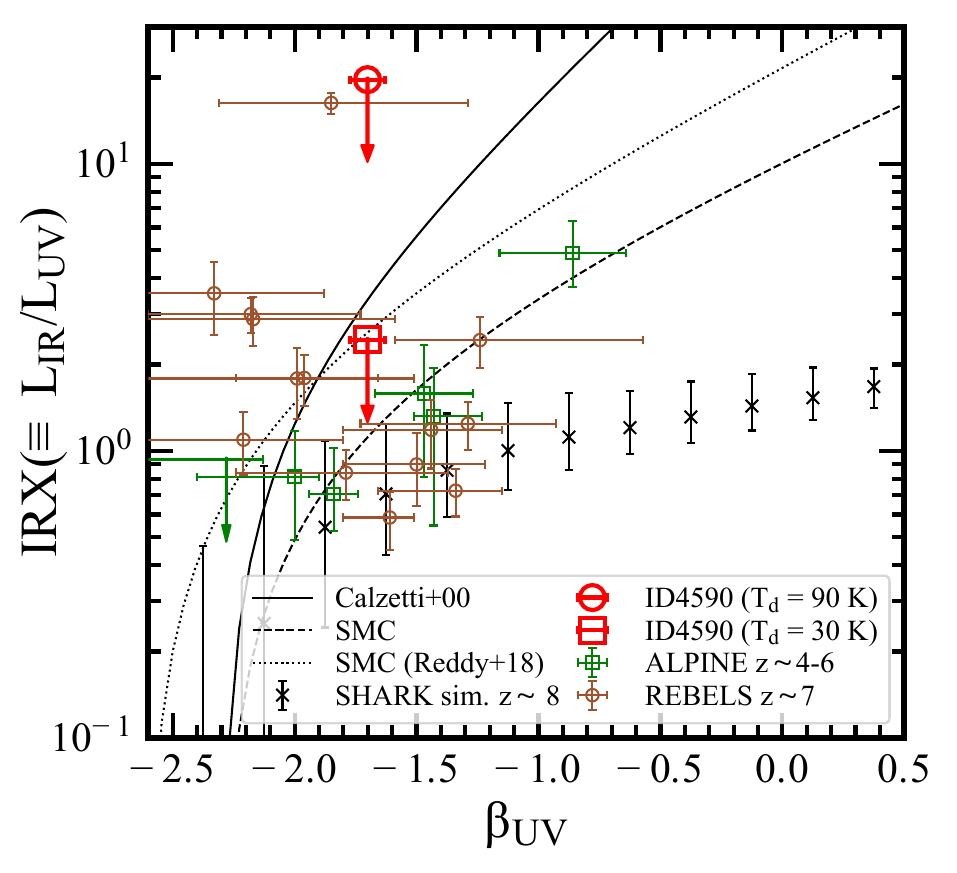}
 \caption{
IRX--$\beta_{\rm UV}$ relation.  
The red symbols represent \targ, where the square and the circle show the upper limits with the $T_{\rm d}$ assumptions of 30~K and 90~K, respectively. 
The solid and dashed curves indicate the relations derived with the dust attenuation of SMC and \cite{calzetti2000}, respectively. 
The dotted curve shows the relation derived from SMC dust attenuation and bluer intrinsic $\beta_{\rm UV}$ \citep{reddy2018}. 
For comparison, we also show other star-forming galaxy results at $z\sim$4--7 taken from ALPINE \citep[green squares; ][]{lefevre2020, fudamoto2020} and REBELS \citep[brown circles;][]{bouwens2021, inami2022}. 
The black crosses indicate semi-analytical simulation results of SHARK \citep{lagos2020}, for galaxies at $z\sim8$ with $\log(M_{\rm star}/M_{\rm star})=7.5$--8.5, where the error bars are the 16--84th percentile. 
\label{fig:irx}}
\end{figure}

Figure \ref{fig:irx} presents the infrared excess ($L_{\rm IR}/L_{\rm UV}\equiv$ IRX) and the UV continuum slope $\beta_{\rm UV}$ relation of \targ. Although we adopt the fiducial value of $T_{\rm d}=60$~K in Section \ref{sec:fir_sed}, 
here we show two extreme cases for \targ, with $T_{\rm d}=$30~K and 90~K, by assuming the potential uncertainty of $\Delta T_{\rm d}=\pm30$~K. 
For comparison, we also show the relations based on several dust attenuation laws \citep{reddy2018, calzetti2000}, a semi-analytical model of SHARK\footnote{
The dust extinction and re-emission are modeled with \cite{charlot2000} and the templates of \cite{dale2014}, respectively, while the final IRX--$\beta_{\rm UV}$ relation is also regulated by dust surface density, $Z_{\rm gas}$, $M_{\rm gas}$, and galaxy structure. 
} 
for galaxies at $z\sim8$ \citep{lagos2020},  and the recent measurements for $z\sim$4--7 star-forming galaxies observed in the ALPINE \citep{lefevre2020} and REBELS surveys \citep{bouwens2021} taken from \cite{fudamoto2021} and \cite{inami2022}, respectively. 
While several Balmer emission lines are detected in \targ, we note that concluding which dust attenuation law fits the best with \targ\ is still challenging with their current S/N and the potential difficulty of the proper aperture correction for each Balmer emission (Section \ref{sec:full_sed}). 

In Figure \ref{fig:irx}, \targ\ falls in a moderately red UV color regime, where the dust continuum is detected from the previous ALMA observations for the $z\sim$ 4--7 galaxies.
If \targ\ has a relatively low $T_{\rm d}$ ($\sim$30--40~K) and follows a dust attenuation law similar to \cite{calzetti2000} or SMC in \cite{reddy2018}, 
the upper limit suggests that the dust continuum should be detected from \targ\ with the current ALMA depth\footnote{
In the energy balance between the dust attenuation and re-emission, a lower $T_{\rm d}$ makes the FIR SED peak wavelength longer at a given $L_{\rm IR}$, where the ALMA sensitivity limit may reach the FIR SED peak at some point.}. 
This indicates that \targ\ has a $T_{\rm d}$ value higher than $\sim$30--40~K or an SMC-like steep dust attenuation law with the intrinsic UV continuum slope of $\beta_{\rm UV,0}\simeq-2.3$ \citep[e.g.,][]{meurer1999, mclure2018}. 
Given the low dust content implied from the low metallicity of \targ\ from the NIRSpec results (Section \ref{sec:jwst}) and the high $\Sigma_{\rm SFR}$ from the compact rest-frame UV size (Section \ref{sec:size}), the dust is efficiently heated at a given UV field \citep[e.g.,][]{behrens2018, sommovigo2022, sommovigo2022b}, and the former high $T_{\rm d}$ scenario might be plausible. 
We also find that the semi-analytical model of SHARK predicts the IRX--$\beta_{\rm UV}$ relation even lower than the SMC relation for the simulated galaxies with similar redshift and $M_{\rm star}$ as \targ, the non-detection of the dust could be simply because of  the difference of the dust attenuation law between local and high-$z$ galaxies. 

Future ALMA high-frequency follow-up observations (e.g., Band~8, 9, and 10) will confirm or rule out the high $T_{\rm d}$ scenario in \targ. 
Furthermore, the detection of the rest-frame UV--optical continuum and/or multiple high-significance detection of Balmer and Paschen emission lines in upcoming \jwst/NIRSpec and MIRI observations will help us to directly constrain the dust attenuation curve in high-redshift galaxies, including \targ. 

\subsection{Onset of outflow at $z>8.5$, facilitating reionization}
\label{sec:extended}

In Figure \ref{fig:nircam}, we find the spatial offset of \cii\ from the \jwst\ source position (Section \ref{sec:morph}). 
The spatial offset suggests the presence of 
the accreting/satellite gas clump(s) \citep[e.g.,][]{maiolino2015} or the extended \cii\ gas beyond the stellar distribution \citep[e.g.,][]{fujimoto2019, fujimoto2020b, ginolfi2020, carniani2020, herrera-camus2021, akins2022, lambert2022}, tracing the diffuse neutral hydrogen \citep{heintz2021, heintz2022}. 
The \cii\ size measurement result, almost $\sim10$ times larger than the stellar distribution of \targ\ (Section \ref{sec:size}), supports the latter scenario of the presence of the extended \cii\ gas. 
Although we cannot rule out the possibility that the noise fluctuation makes the \cii\ morphology look extended with the current S/N, it is worth mentioning that similarly extended \cii\ morphology with spatial offsets have also been observed in other galaxies at $z\sim7$--9 \citep{carniani2020}. 
The other possibility could be nearby faint, dusty objects \citep[e.g.,][]{fujimoto2016, fujimoto2022} emitting \cii\ \citep[e.g.,][]{romano2020, fudamoto2022}. 
However, no counterparts are identified down to a 3$\sigma$ upper limit in F150W of $\sim$32.5 mag (with $0\farcs2$-diameter; \citealt{harikane2022b}) after the lens correction, which corresponds to SFR $\approx0.05\,M_{\odot}$~yr$^{-1}$ at $z=8.5$ \citep[e.g.,][]{kennicutt2012}. 
Similarly, no counterparts in F444W place a $3\sigma$ upper limit of $M_{\rm star}\lesssim1\times10^{6}\,M_{\odot}$ by scaling the best-fit SED of \targ. 
A pixel-based SED analysis in \cite{clara2022b} shows the gradient of the dust obscuration decreasing towards the \cii-emitting region, which also supports the absence of the counterparts at the \cii\ peak position.

Whichever the \cii-emitting gas is compact clump(s) diffuse \& extended, these results suggest that the carbon in the gas is illuminated not by local star-forming activities, but by i) the ionizing photon escaped from inside the galaxy, ii) shock heating of the outflowing gas, or iii) a cooling process of the hot outflowing gas \citep[see also discussions in e.g.,][]{fujimoto2019, fujimoto2020b, pizzati2020, akins2022}. 
In Figure \ref{fig:sfr_line}, we find that the $L_{\rm [CII]}$/SFR ratio of this \cii\ emission outside of galaxy is consistent with other local dwarf galaxies whose $Z_{\rm gas}$ values are similarly low as \targ\ (IZw18 and SBS0335-052). 
This suggests that the input energy from i) is enough to explain the observed $L_{\rm [CII]}$. 
In fact, the high ionization parameter of $\log(U)\gtrsim-2$ observed in \targ\ is in line with the scenario of i). 

In Figure \ref{fig:nircam}, we also find the extended ionized gas (\oiii$\lambda$5007, H$\beta$) structure in the deep NIRCam/F444W filter (Section \ref{sec:morph}). 
Similar to the \cii\ emission, no NIRCam counterparts are identified in the extended ionized gas regions. 
In addition, the extended ionized gas regions are also not matched with the direction of the high dust obscuration gradient, which denies the possibility that the obscured star-forming regions cause the extended ionized emission. 
Based on the size measurement results in the image plane (Section \ref{sec:size}) and the different radial and tangential magnifications (see middle panel of Figure \ref{fig:nircam}), the ionized \oiii$\lambda$5007+H$\beta$ structure extends out to $>8$ times more than the effective radius of the stellar distribution of \targ\ (Section \ref{sec:morph}). 
Same as \cii, i)--iii) are the possible physical origins of the extended ionized gas emission. Interestingly, from the center of \targ, the regions of the extended ionized gas distribution ($\sim$Southeastern) and the peak position of the \cii\ emission ($\sim$Northern) are in different directions. This might indicate that the differential distributions of the nebular parameters, such as $n_{\rm e}$, $Z_{\rm gas}$, $\log(U)$, and C/O abundance cause these differential distributions, even if the same physical process is taking place, either of i)---iii). 
The different morphology between \oiii\~88$\mu$m and \oiii5007+H$\beta$ might also be attributed to the differential distributions of the nebular parameters, although the insufficient depth in the ALMA 88$\mu$m map could be another plausible cause.

Importantly, regardless of the true physical processes to give rise to an offset-\cii\ and extended \oiii$\lambda$5007+H$\beta$ gas emission, the presence of the metal-enriched gas away from the galaxy is strong evidence of past and/or ongoing outflow activities already taking place in a low-mass ($M_{\rm star}=6\times10^{7}\,M_{\odot}$), metal-poor ($Z=0.04Z_{\odot}$) nascent galaxy at $z=8.496$. 
The presence of the extended carbon gas structure, so-called \cii\ Halo, has been reported around more massive ($M_{\rm star}>10^{9}$--10$^{10}\,M_{\odot}$) star-forming galaxies at $z=4-7$ \citep[e.g.,][]{fujimoto2019, fujimoto2020b, ginolfi2020, herrera-camus2021, akins2022, lambert2022}, which is challenging to the current cosmological galaxy formation models (\citealt{fujimoto2019}; see also e.g., \citealt{pizzati2020, arata2020, katz2022}). 
Our results in \targ\ provide new insight that such metal enrichment beyond the galaxy ISM scale starts to occur even in the very early phase of the galaxy assembly just $\sim$580~Myr after the Big Bang, which is likely linked to the origin of the \cii\ halo at a later epoch of the Universe. 

Another important fact is that the presence of diffuse extended ionized gas around \targ\ directly indicates the high filling factor of the ionized gas, where the ionizing photons escaping from the galaxy may contribute to the Reionization. 
After the lens correction, \targ\ is $\sim5\times$ fainter than the characteristic UV luminosity of the UV luminosity function (UVLF) at $z\sim9$ ($M^{*}_{\rm UV}=-19.6$ mag; e.g., \citealt{harikane2022b}). 
Once we confirm the high escape fraction of the ionizing photons from faint, low-mass galaxies, probably related to the onset of the outflow from the early stage of the galaxy assembly, it also provides us with a new insight into the process of the Reionization, in contrast to the scenario that huge ionized bubbles are formed around UV-bright galaxies ($M_{\rm UV}\lesssim-22$) at similar redshifts \citep[e.g.,][]{mason2018}.
The significantly low metallicity of \targ, falling below the $z\sim8$ mass--metallicity relations predicted from current galaxy formation models \citep[e.g.,][]{curti2022}, 
despite the dust obscuration in \targ\ ($\beta=-1.7\pm0.07$), 
may suggest that dust obscuration occurs in a part of the galaxy with a very low dust content. 
This may also be caused by the past or ongoing outflow activities \citep{ferrara2022,ziparo2022} that carry the dust away from the galaxy and make it diffuse, cold, and undetectable in the observations \citep[e.g.,][]{akins2022}, while the dust in regions that are not invested by the outflow, or that are shielded by Giant Molecular Clouds, could survive (see also \citealt{martinez-gonzalez2019, nath2022}). 
This small amount of surviving dust may be responsible for a certain amount of dust obscuration.
In any case, the low dust content in the galaxy is also helpful for ionizing photons to escape from the system.

Note that diffuse ionized gases (DIGs) have been observed in local galaxies, from inter-arm regions \citep[e.g.,][]{zurita2000} to areas above the galactic mid-plane out to 1--2~kpc scales \citep[e.g.,][]{rossa2000}. However, even in the latter case, the sizes of these DIGs are only about $\sim10$\% relative to the size of the host galaxy \citep[e.g.,][]{rossa2003b}. In contrast, the extended ionized gas structure around \targ\ is well beyond the central galaxy size (Section \ref{sec:morph}). Therefore, the physical origins of the extended ionized gas structure around \targ\ are likely different from those of the DIGs in the local Universe. 

Another note is that we do not find evidence of the ongoing outflow via the broad-wing feature in the NIRSpec spectrum. However, the slit of the NIRSpec MSA is aligned perpendicular to the extended ionized gas structure (See the white rectangle in Figure \ref{fig:nircam}), which might be the reason for the absence of the broad-wing feature in the current NIRSpec spectrum. 

\begin{figure*}
\includegraphics[trim=0cm 0cm 0cm 0cm, clip, angle=0,width=1\textwidth]{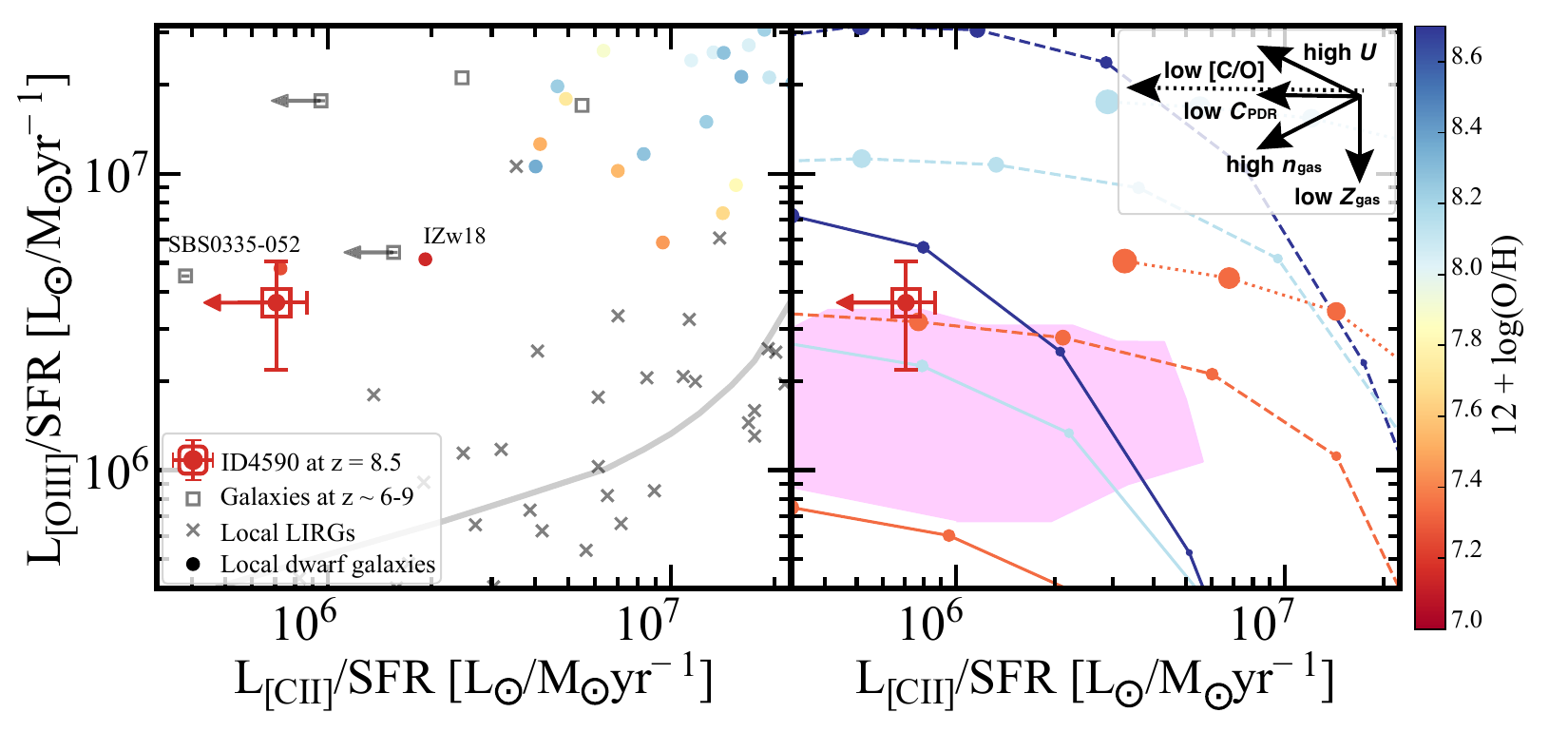}
 \caption{
FIR line diagnostic of $L_{\rm [OIII]}$/SFR and $L_{\rm [CII]}$/SFR, produced in the same manner as Figure 12 in \cite{harikane2020}.  
\textbf{\textit{Left:}}
The relation from the observations. 
The color and symbols represent the same as Figure \ref{fig:sfr_line}. 
The grey crosses are newly added in this panel, showing the local LIRGs \citep{howell2010, diaz-santos2017} whose SFR values are comparable to $z\sim6$--9 galaxies in the panel. 
The grey curve is the typical relation for the local LIRGs. 
\textbf{\textit{Right:}}
Same as the left panel, but comparing with {\sc Cloudy} calculations. 
The magenta-shaded region represents the possible parameter space for \targ\ calcualted with {\sc Cloudy} based on its best-fit nebular parameters of $Z_{\rm gas}=0.04\pm0.02$, $n_{\rm e}=220^{+230}_{-130}$~cm$^{-3}$, $\log(U)>-2.27$, and $\log$(C/O)=$[-0.52:-0.26]$ constrained by our joint \jwst\ and ALMA analysis. 
For comparison, we also show the relations with different parameter sets;  
the orange, dark light blue, and dark blue lines are the results with $Z_{\rm gas}=0.05, 0.1,$ and 1.0~$Z_{\odot}$; the dotted, dashed, and solid lines correspond to densities of $\log(n_{\rm gas}$/[cm$^{-3}$]) = 0.5, 2.0, and 3.0; and the larger circles indicate higher ionization parameters, from $\log U$ = $-4.0$ ion to $-0.5$ with a step size of 0.5.
The black arrows in the inset panel show directions and possible shifts in the $L_{\rm [OIII]}/$SFR--$L_{\rm [CII]}/$SFR plane due to the physical mechanisms of  b), c), and d).
To perform a fair comparison, we use the SFR value based on the dust-corrected H$\alpha$ luminosity in the same manner as \cite{harikane2020}, instead of the SED based estimate. Using H$\alpha$ luminosity is also beneficial to count the ionizing photons contributing to the line emissivities more directly. 
\label{fig:cloudy}}
\end{figure*}

\subsection{Physical origins of the high [OIII]88$\mu$m/[CII]158$\mu$m ratio} 
\label{sec:oiii-cii_ratio}

Previous ALMA observations reported the detections of the luminous \oiii88$\mu$m line from star-forming galaxies at $z\sim$ 6--9, showing their \oiii88$\mu$m/\cii158$\mu$m line ratio of $\gtrsim3$--10 that are higher than local dwarf galaxies and/or local Luminous Infrared Galaxies (LIRGs) with similar SFRs \citep{inoue2016, harikane2020, witstok2022}.
The origin of such high ratios is still unclear. Major solutions that have been argued include 
a) observational bias  \citep[e.g.,][]{carniani2020}, 
b) low C/O abundance ratio \citep{katz2022}, 
c) low covering fraction of PDRs\footnote{
This is equal to the high filling factor of the ionized gas relative to dense PDRs. 
} 
$C_{\rm PDR}$ \citep{cormier2015, harikane2020}, 
and d) the characteristic ISM parameters in early galaxies such as the high ionization parameter \citep{katz2017, harikane2020} probably caused by recent strong bursts of star-formation \citep{ferrara2019, arata2020, vallini2021, witstok2022, sugahara2022} and/or low stellar-to-gaseous metallicity ratio \citep{sugahara2022}. 

In Section \ref{fig:sfr_line}, we find that \targ\ also shows a similarly high \oiii88$\mu$m/\cii158$\mu$m line ratio of $>4$ at the galaxy position.  
In addition to the accurate measures of $Z_{\rm gas}$ and $n_{\rm e}$ (Section \ref{sec:line_ratio}), the deep NIRSpec observations also cover the \oii3729 and \ciii1909 emission lines from \targ, which allow us to constrain the ionization parameter $\log(U)$ (e.g., \citealt{schaerer2022, curti2022, brinchmann2022, trump2022}; \citealt{heintz2022c}; \citealt{nakajima2023}) and the C/O abundance in \targ\ (e.g., \citealt{arellano2022}; \citealt{isobe2023b}). 
This indicates that we can investigate the physical origins of the high \oiii88$\mu$m/\cii158$\mu$m line ratio, taking the possible solutions of b) and d) into account via the actual observed measurements for \targ. 
Besides, the spatial offset of the \cii\ line (Section \ref{sec:morph}) allows us to separate the emission arising inside and outside the galaxy and fairly compare the line ratio, also managing the observational bias of point a). 
Therefore, with this best optical--mm characterization of an early galaxy owing to the joint \jwst\ and ALMA analysis, we are ready to address the physical origins of the high \cii/\oiii\ and verify whether the remaining possible solution of c) is critical or other solutions can answer already. 

In Figure \ref{fig:cloudy}, we show the $L_{\rm [OIII]}/$SFR and $L_{\rm [CII]}/$SFR relation of \targ. 
Given the purpose of the analysis, we use the \cii\ results at the galaxy position. 
For comparison, we also present the relations from observations (left panel) and the predictions from the photoionization model with {\sc cloudy} (right panel) drawn in the same manner as \cite{harikane2020} (see also \citealt{sugahara2022}), where the possible shifts on the plane by b), c), and d) are presented with black arrows \citep{harikane2020}. 
We also show the magenta-shaded region that corresponds to the possible space for \targ\ calculated by {\sc Cloudy} with our fiducial estimates of $Z_{\rm gas}/Z_{\odot}=0.04\pm0.02$, $n_{\rm e}=220^{+230}_{-130}$~cm$^{-3}$, $\log(U)>-2.27$, and $\log$(C/O)=$[-0.52:-0.26]$. 
Following \cite{harikane2020}, we use the SFR value based on the dust-corrected H$\alpha$ luminosity estimated from NIRSpec (Section \ref{sec:jwst}), instead of the SED-based value, for this analysis. 

In the left panel, we find that \targ\ shows the $L_{\rm [OIII]}$/SFR ratio much higher than local LIRGs, falling on the high \oiii/\cii\ line ratio regime similar to other $z\sim6$--9 galaxies. 
We also find that several metal-poor galaxies (12+$\log(O/H)\lesssim8.0$) among the local dwarf galaxies are located in the similarly high \oiii/\cii\ line ratio regime to these $z\sim6$--9 galaxies. This may suggest that the ISM conditions of these $z\sim6$--9 galaxies are similar to those of the local metal-poor galaxies, while an important note is that these local metal-poor galaxies have much lower SFRs than these $z\sim6$--9 galaxies by $\sim$1--2 orders of magnitudes.

In the right panel, we find that the observed $L_{\rm [OIII]}$/SFR and $L_{\rm [CII]}$/SFR relation of \targ\ is consistent with the magenta-shaded region within the errors. 
This indicates that the \cii\ and \oiii\ emissivities at a given input energy in \targ\ are generally explained by the combination of high $\log(U)$, high $n_{\rm e}$, low $Z_{\rm gas}$, and low $\log$(C/O). 
Therefore, the physical origin of the high \oiii/\cii\ ratio observed among $z\sim6$--9 galaxies may be sufficiently explained by b) and d).

As discussed in \cite{harikane2020}, another possible origin of c) -- the low $C_{\rm PDR}$ -- could also be the additional reason to boost the \oiii/\cii\ ratio. As indicated in the black arrow, this effect makes the data points horizontally to move to the left in Figure \ref{fig:cloudy} (i.e., towards low \cii/SFR). 
Because the current $\log(U)$ estimate provides the upper limit alone, 
the possible parameter space of \targ\ extends to the low $L_{\rm [CII]}$/SFR regime, 
where we cannot disentangle the contributions from $\log(U)$ and $C_{\rm PDR}$. 
Once the upper boundary of $\log(U)$ is constrained, the lower limit of the $L_{\rm [CII]}$/SFR ratio in the possible parameter space will be determined, where the additional $C_{\rm PDR}$ contribution will be evaluated if the observed $L_{\rm [CII]}$/SFR upper limit is lower than the possible parameter space. 
With future deep \cii\ follow providing a much more stringent upper limit or the faint detection of \cii, we may be able to investigate further the additional contribution from $C_{\rm PDR}$.

\section{Summary}
\label{sec:summary}
In this paper, we present the ALMA Band~7 and Band~5 deep spectroscopy for the two major coolant FIR lines of \oiii88~$\mu$m and \cii158~$\mu$m from \targ, a metal-poor, low-mass, strongly lensed sub-$L^{*}$ galaxy at $z=8.496$, whose warm interstellar medium and stellar properties have been the best characterized with the \jwst\ Early Release Observations (ERO) for the SMACS0723 field. 
The \jwst\ ERO observations include the deep imaging with NIRCam and MIRI at $\sim$1--20~$\mu$m as well as deep spectroscopy with NIRSpec at $\sim$2--5~$\mu$m which detects multiple rest-frame optical emission lines, including \oiii$\lambda$4363 line, and provides us with the robust measure of the gas-phase metallicity via the \textit{direct} temperature method for the first time at $z\gtrsim3$. 
Combining these rich \jwst\ data with \hst\ and ALMA, the high-spatial-resolution, homogeneous data set from optical to mm wavelengths enables us to perform a panchromatic characterization of an early galaxy in and out, which sets the benchmark for synergetic studies of ALMA and \jwst\ in the coming decades. 
The main findings of this paper are summarized as follows:
\begin{enumerate}
\item We detect both \cii158$\mu$m and \oiii88$\mu$m lines at 4.0$\sigma$ and 4.5$\sigma$ levels, respectively. The redshift and the spatial position of the \oiii\ line exactly coincide with those of the \jwst\ source. On the other hand, the \cii\ line is blue-shifted by 90~km~$^{-1}$ and has a spatial offset by $\sim0\farcs5$ ($\approx0.5$~kpc in source plane) beyond the errors, where the chance projection of the noise fluctuation is very unlikely ($\sim0.07$\%) from a blind line search analysis in the same data cube. This indicates that \cii\ velocity and spatial offsets are real and associated with some physical mechanisms of \targ. 
\item We evaluate the sizes of \cii, \oiii, rest-frame UV, and rest-frame optical continuum by using ALMA and \jwst/NIRCam images. With the 2D profile fitting in the image plane, we obtain the best-fit circularized effective radii of $0\farcs69\pm0\farcs42$, $0\farcs056\pm0\farcs012$, and $0\farcs059\pm0\farcs010$ for \cii, rest-frame UV, and rest-frame optical continuum, respectively. The \oiii\ line is not spatially resolved with the current resolution and sensitivity, and we place its upper limit of $0\farcs16$. 
We find that the position angle estimated in the F150W filter shows an excellent agreement with the lensing distortion predicted from the mass model, validating the high magnification of $\mu=8.69$ in \targ.  
The extended \cii\ distribution relative to the stellar continuum is reminiscent of the \cii\ Halo reported in recent ALMA studies, while we cannot exclude the possibility that the \cii\ size estimate is affected by the noise fluctuation with the current S/N. 
\item The dust continuum is not detected in neither Band~7 or Band~5, and we place 2$\sigma$ upper limits of 41.8~$\mu$Jy and 23.2~$\mu$Jy at 0.85~mm and 1.54~mm, respectively. By assuming a single modified black body with the dust temperature $T_{\rm d}=60$~K and the dust spectral index of $\beta_{\rm d}=1.8$, we estimate the upper limits of infrared luminosity over 8--1000~$\mu$m of $L_{\rm IR}<1.8\times10^{10}\,L_{\odot}$ and the dust mass of $M_{\rm dust}<1.2\times10^{5}\,M_{\odot}$ after the lens correction, while a different $T_{\rm d}$ assumption by $\pm30$~K changes these constraints by $\sim\pm$0.5--1.0~dex. 
The upper limit of infrared excess (IRX$\equiv L_{\rm IR}/L_{\rm UV}$) in \targ\ suggests that $T_{\rm d}$ is higher than $30$--40~K or that steep dust attenuation laws similar to SMC are favored. 
\item The NIRCam/F444W filter, including the contributions from \oiii$\lambda$5007 and H$\beta$ emission lines, shows an extended structure where the rest-frame UV and optical continuum are invisible and the 3$\sigma$ upper limit in the rest-frame UV continuum after the lens correction are placed at 32.5~mag ($\approx0.05\, M_{\odot}$~yr$^{-1}$). Given the high-spatial-resolution of the F444W filter ($\sim0\farcs15$), the smooth morphology of the extended structure is the direct evidence of the presence of the extended, diffuse ionized gas structure around \targ. The structure extends towards the radial magnification axis out to $\sim0\farcs5$. After the lens correction, this corresponds to $\sim1.7$~kpc and at least eight times larger than the rest-frame optical effective radius of \targ.  
\item We perform the optical--mm SED analysis with \hst, JWST/NIRCam+MIRI, and ALMA photometry. We exclude the emission outside of the galaxy observed in the F444W filter, and the remaining contribution of the strong emission lines of \oiii$\lambda$5007+H$\beta$ and the stellar continuum is well separated by the rich filter sets of NIRCam and MIRI. After the lens correction, we estimate the stellar mass of $M_{\rm star}=6\times10^{7}\,M_{\odot}$, the total star-formation of SFR $=3\,M_{\odot}$~yr$^{-1}$, UV continuum slope of $\beta_{\rm UV}=-1.7$, suggesting that \targ\ is a low-mass, but a little dust attenuated galaxy, in contrast to very blue galaxies ($\beta_{\rm UV}<-2.0$) that have been observed in recent \jwst\ observations at similar redshifts.  
\item Regardless of the ongoing physical mechanisms, past outflow activities are required to make the surrounding pristine gas of \targ\ metal-enriched and produce the \cii\ offset and the extended ionized gas structure traced by \oiii$\lambda$5007+H$\beta$. This would also help produce high ionizing photon escape from \targ\ and contribute to reionization at $z>8.5$. 
\item With careful slit-loss correction and the separation of the emission inside and outside of the galaxy, we evaluate the electron density of $n_{\rm e}=220^{+230}_{-130}$~cm$^{-3}$ via the \oiii88$\mu$m/\oiii$\lambda$5007 line ratio. This is much higher than that of local galaxies ($n_{\rm e}\simeq30$~cm$^{-3}$), but consistent with $z\sim2$--3 galaxies ($n_{\rm e}\simeq200-300$~cm$^{-3}$). This is also consistent with the upper limit of $n_{\rm e}<260$~cm$^{-3}$ obtained in a lensed dusty galaxy at $z=7.13$. 
\item We examine relations between the line luminosities of \cii158$\mu$m, \oiii88$\mu$m, and SFR. \targ\ shows $L_{\rm [CII]}$--SFR and $L_{\rm [OIII]}$--SFR relations generally consistent with other $z\sim6$--9 galaxies and explores the faint-end of the relations, owing to the aid of the lensing support.  
The $L_{\rm [CII]}$/SFR ratio of \targ\ falls below the typical relation estimated among the local dwarf galaxies beyond the errors. Still, it is consistent with similarly metal-poor local galaxies of IZw18 and SBS0335-052. The same result is obtained in the SFR--$L_{\rm [OIII]}$ relation, while the relation of \targ\ is still consistent with the typical range of the local dwarf galaxies within the errors. 
\targ\ shows a $L_{\rm [OIII]}/L_{\rm [CII]}$ ratio of $>4$, which is also as high as other $z\sim$6--9 galaxies, falling above the typical relation of the local dwarf galaxies in the SFR--$L_{\rm [OIII]}/L_{\rm [CII]}$ relation. 
\item We investigate the physical origins of the high $L_{\rm [OIII]}/L_{\rm [CII]}$ ratio with the photoionization model of {\sc Cloudy}. The $L_{\rm [OIII]}$/SFR--$L_{\rm [CII]}$/SFR relation of \targ\ is generally reproduced by the high $n_{\rm e}$, low gas-phase metallicity ($Z_{\rm gas}/Z_{\odot}=0.04$), high ionization parameter ($\log(U)>-2.27$), and low carbon-to-oxygen abundance ratio $\log$(C/O)$=[-0.52:-0.24]$ obtained from the \jwst/NIRSpec data. 
While the other potential mechanism of the low covering fraction of the photodissociation region is not constrained by the current data, it will be achieved by further deep ALMA \cii\ follow-up. 
\end{enumerate}

We thank the anonymous referee for constructive comments and suggestions.
We also thank Catherine Vlahakis and Holly Sheets for their kind support in the processing of the ALMA data, despite the unusual circumstances due to the cyberattack.
We are grateful to Pablo Albalo Halo for sharing deep insights into the NIRSpec data analysis and to Caitlin Casey, Steven Finkelstein, and Daniel Stark for helpful discussions for this paper. 
This paper makes use of the ALMA data: ADS/JAO. ALMA \#2022.A.00022.S. 
ALMA is a partnership of the ESO (representing its member states), 
NSF (USA) and NINS (Japan), together with NRC (Canada), MOST and ASIAA (Taiwan), and KASI (Republic of Korea), 
in cooperation with the Republic of Chile. 
The Joint ALMA Observatory is operated by the ESO, AUI/NRAO, and NAOJ. 
This work is based on observations and archival data made with the {\it Spitzer Space Telescope}, which is operated by the Jet Propulsion
Laboratory, California Institute of Technology, under a contract with NASA, along with archival data from the NASA/ESA 
{\it Hubble Space Telescope}. 
This research also made use of the NASA/IPAC Infrared Science Archive (IRSA), 
which is operated by the Jet Propulsion Laboratory, California Institute of Technology, under contract with the National Aeronautics and Space Administration. 
The Early Release Observations\footnote{
\url{https://www.stsci.edu/jwst/science-execution/approved-programs/webb-first-image-observations}} and associated materials were developed, executed, and compiled by the ERO production team:  Hannah Braun, Claire Blome, Matthew Brown, Margaret Carruthers, Dan Coe, Joseph DePasquale, Nestor Espinoza, Macarena Garcia Marin, Karl Gordon, Alaina Henry, Leah Hustak, Andi James, Ann Jenkins, Anton Koekemoer, Stephanie LaMassa, David Law, Alexandra Lockwood, Amaya Moro-Martin, Susan Mullally, Alyssa Pagan, Dani Player, Klaus Pontoppidan, Charles Proffitt, Christine Pulliam, Leah Ramsay, Swara Ravindranath, Neill Reid, Massimo Robberto, Elena Sabbi, Leonardo Ubeda. The EROs were also made possible by the foundational efforts and support from the JWST instruments, STScI planning and scheduling, and Data Management teams.
This project has received funding from the European Union’s Horizon 2020 research and innovation program under the Marie Sklodowska-Curie grant agreement No. 847523 ‘INTERACTIONS’ and from NASA through the NASA Hubble Fellowship grant HST-HF2-51505.001-A awarded by the Space Telescope Science Institute, which is operated by the Association of Universities for Research in Astronomy, Incorporated, under NASA contract NAS5-26555.
The project leading to this publication has received support from ORP, that is funded by the European Union’s Horizon 2020 research and innovation programme under grant agreement No 101004719 [ORP]. 
F.S. acknowledges support from the NRAO Student Observing Support (SOS) award SOSPA7-022.
The Cosmic Dawn Center is funded by the Danish National Research Foundation under grant No. 140.
K.~Kohno acknowledges the support by JSPS KAKENHI Grant Number 17H06130 and the NAOJ ALMA Scientific Research Grant Number 2017-06B.
The National Radio Astronomy Observatory is a facility of the National Science Foundation operated under cooperative agreement by Associated Universities, Inc.

\software{{\sc casa} (v6.4.1; \citealt{casa2022}), 
\texttt{grizli} \citep{brammer2023}, 
\texttt{EAZY}, 
\citep{brammer2008}, 
\texttt{PyNeb} \citep{shaw1998}
\texttt{CLOUDY}, 
\citep{ferland2017}, 
\texttt{Interferopy}, \citep{boogaard2021}, 
and \texttt{CIGALE} \citep{boquien2019}}. 

\clearpage

\appendix

\section{Accuracy of the astrometry of \jwst\ and \hst\ maps from grizli}

\begin{figure}
\includegraphics[trim=0cm 0cm 0cm 0cm, clip, angle=0,width=0.45\textwidth]{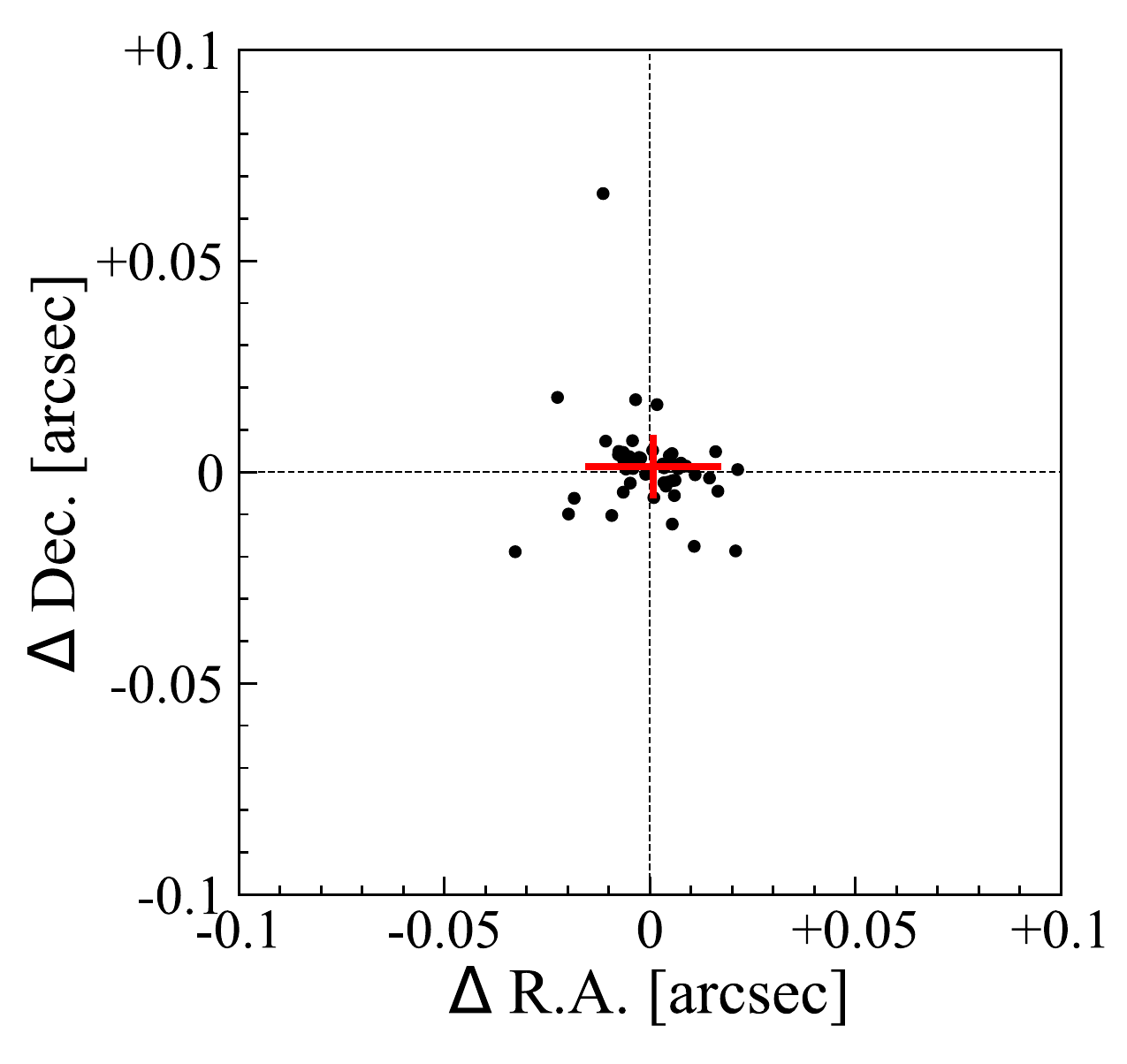}
 \caption{
Accuracy of astrometry of the \jwst\ and \hst\ maps used in our analysis. The black circles show the relative offsets of the coordinates from the GAIA DR3 catalog for the GAIA sources identified in the \jwst\ maps, after correcting their proper motion effects. 
The red cross indicates the average offset, showing excellent agreement with zero. 
The red cross size corresponds to the standard deviations of $\Delta$R.A. and $\Delta$Dec., which we assume as the uncertainty of the positional accuracy for the \jwst\ sources (black bar in Figure \ref{fig:nircam}).   
\label{fig:astrometry}}
\end{figure}

In Figure \ref{fig:astrometry}, we show the residual spatial offsets of the Right ascension ($\Delta$R.A.) and Declination ($\Delta$Dec.) for the GAIA sources. We evaluate the residual offsets by running the {\sc source extractor} \citep{bertin1996} on the \jwst\ and \hst\ maps produced by the \texttt{grizli} pipeline and being corrected for the proper motion effect of the GAIA sources, and comparing with those from the GAIA DR3 catalogs. 
We confirm that the average offset shows an excellent agreement with zero and the standard deviation in R.A and Dec. of $0\farcs01$ and $0\farcs02$, respectively. We regard these standard deviations as the positional uncertainty of the \jwst\ sources in Figure \ref{fig:nircam}. 

\section{CIGALE parameters for the full SED fit}

We assume a delayed star-formation history (SFH): SFR($t$) $\propto t/\tau^{2}$ exp$(-t/\tau)$ with stellar models from \cite{bruzual2003} and an exponential recent burst implemented in the \texttt{sfhdeleyed} module. 
We use the \texttt{dustatt\_modified\_starburst} module for the dust attenuation, where the nebular emission (continuum + lines) are attenuated with a screen model and an SMC extinction curve \citep{pei1992}. During the SED fitting, the same $E(B-V)$ is used between stellar and nebular emission. Finally, the dust emission is re-emitted in the infrared with \cite{casey2012} models.
We fix $\beta_{\rm d}=1.8$ and the mid-infrared power-law slope $\alpha_{\rm d}=2.0$ and adopt the $T_{\rm d}$ range of 30--90~K.
Because no evidence of the presence of AGN has been reported, we do not include an AGN component in the model. 
We list the modules and the parameter ranges used in the fitting in Table \ref{Tab.CIGALE.Ph2}. In the fitting, we use the photometry with the $1\sigma$ error also for the measurements below the $2\sigma$ upper limits. To avoid the lens model uncertainty, we perform the SED fitting with the optical--mm photometry without the lens correction.

\begin{table*}[h]
\begin{center}
\resizebox{1.\linewidth}{\height}
{
\begin{tabular}{|>{\centering}p{8.0cm}|>{\centering\arraybackslash}p{3.5cm}|>{\centering\arraybackslash}p{3.5cm}|}
  \hline\hline
  {\bf Parameters} & {\bf Symbol} & {\bf Range} \\
  \hline\hline
\multicolumn{3}{c}{}\\
\multicolumn{3}{c}{\bf Delayed SFH and recent burst}\\
  \hline
 e-folding time scale of the delayed SFH & $\tau_{main}$ [Myr] & 100, 250, 500, 1000 \\
  \hline
 Age of the main population & Age$_{main}$[Myr]  & 51 log values in [1: 3.3] \\
  \hline
 Burst & f$_{burst}$  &  0.05, 0.10, 0.15, 0.25  \\
  \hline
    \multicolumn{3}{c}{}\\
    \multicolumn{3}{c}{\bf SSP}\\
  \hline
  SSP &   & BC03 \\
  \hline
  Initial mass function &  IMF & Chabrier \\
  \hline
  Metallicity     & Z &  0.0004, 0.004, 0.02\\ \hline
    \multicolumn{3}{c}{}\\
    \multicolumn{3}{c}{\bf Nebular emission}\\
  \hline
  Ionization parameter &  logU    & -2.0 \\ \hline
  Line width [km/s]    &     ---    &  150 \\ \hline
  Gas-phase Metallicity &  zgas   & 0.0004, 0.004, 0.02 \\ \hline
  Electron density      & ne      & 100 \\
  \hline
    \multicolumn{3}{c}{}\\
    \multicolumn{3}{c}{\bf Dust attenuation law}\\
  \hline
  Color excess for both the old and young stellar populations &  E\_BV\_lines & 21 log values in [$-3$: 1.3] \\
  \hline
  Reduction factor to apply on E\_BV\_lines to compute E(B-V)s the stellar continuum attenuation & E\_BV\_factor & 1.0 \\ 
  \hline
  Bump amplitude &  uv\_bump\_amplitude &  0.0 \\
  \hline
  Power law slope & power law\_slope &  0.0 \\
  \hline
  Extinction law to use for attenuating the emission lines flux &  Ext\_law\_emission\_lines & SMC \\
  \hline
  Ratio of total to selective extinction, A\_V / E(B-V) & Rv & 2.93 \\
  \hline
    \multicolumn{3}{c}{}\\
    \multicolumn{3}{c}{\bf Dust emission (Casey2012)}\\
  \hline
  Dust temperature & temperature & 20 log values in [1.5:1.95]  \\
  \hline
  Dust emissivity index & beta & 1.8  \\
  \hline
  Mid-infrared power law slope & alpha & 2.0 \\
   \hline
    \hline
\multicolumn{3}{c}{}\\
\multicolumn{3}{c}{\bf No AGN emission}
\\ \hline \hline
\end{tabular}
}
  \caption{{\sc cigale} modules and input parameters used for all the fits. BC03 indicates \cite{bruzual2003}, and the Chabrier IMF refers to \cite{chabrier2003}.}
  \label{Tab.CIGALE.Ph2}
\end{center}
\end{table*}

\section{HST/NIRcam/MIRI photometry of ID4590}
\label{sec:photo}

We summarize the optical--mm photometry used in this paper in Table \ref{tab:photometry}. 

\setlength{\tabcolsep}{20pt}
\begin{table}
\begin{center}
\caption{NIR--mm photometry of ID4590 used in this paper}
\label{tab:photometry}
\begin{tabular}{cc}
\hline 
\hline
Filter & Flux (nJy) \\ \hline
HST/F105W & $-4.7\pm12.3$ \\
HST/F125W & $44.5\pm13.8$ \\
HST/F140W & $60.5\pm11.9$ \\
HST/F160W & $61.7\pm14.7$ \\ \hline
NIRCam/F090W & $-12.0\pm3.9$ \\
NIRCam/F115W & $22.7\pm8.2$ \\
NIRCam/F150W & $60.7\pm6.6$ \\
NIRCam/F200W & $66.9\pm7.0$ \\
NIRCam/F277W & $69.0\pm7.1$ \\
NIRCam/F356W & $71.3\pm7.3$ \\
NIRCam/F444W & $145.2\pm14.7$ \\ \hline
MIRI/F770W & $92.2\pm38.0$ \\
MIRI/F1000W & $111.2\pm72.0$ \\
MIRI/F1500W & $314.1\pm234.4$ \\
MIRI/F1800W & $263.9\pm464.6$ \\ \hline
ALMA/Band7 & $<41800\, (2\sigma)$ \\
ALMA/Band5 & $<23200\, (2\sigma)$ \\ \hline\hline
\end{tabular}
\end{center}
\tablecomments{
We adopt the $0\farcs36$-diameter aperture photometry corrected to the total flux, where we set an error floor of 10\%. 
The ALMA photometry shows the upper limits at the 2$\sigma$ level, assuming that the emission is unresolved with the synthesized ALMA beams (FWHM$\sim0\farcs7$ in Band~7 and $\sim1\farcs3$ in Band~5). 
}
\end{table}

\section{Size measurements}

\begin{figure}[h]
\includegraphics[trim=0cm 0cm 0cm 0cm, clip, angle=0,width=0.5\textwidth]{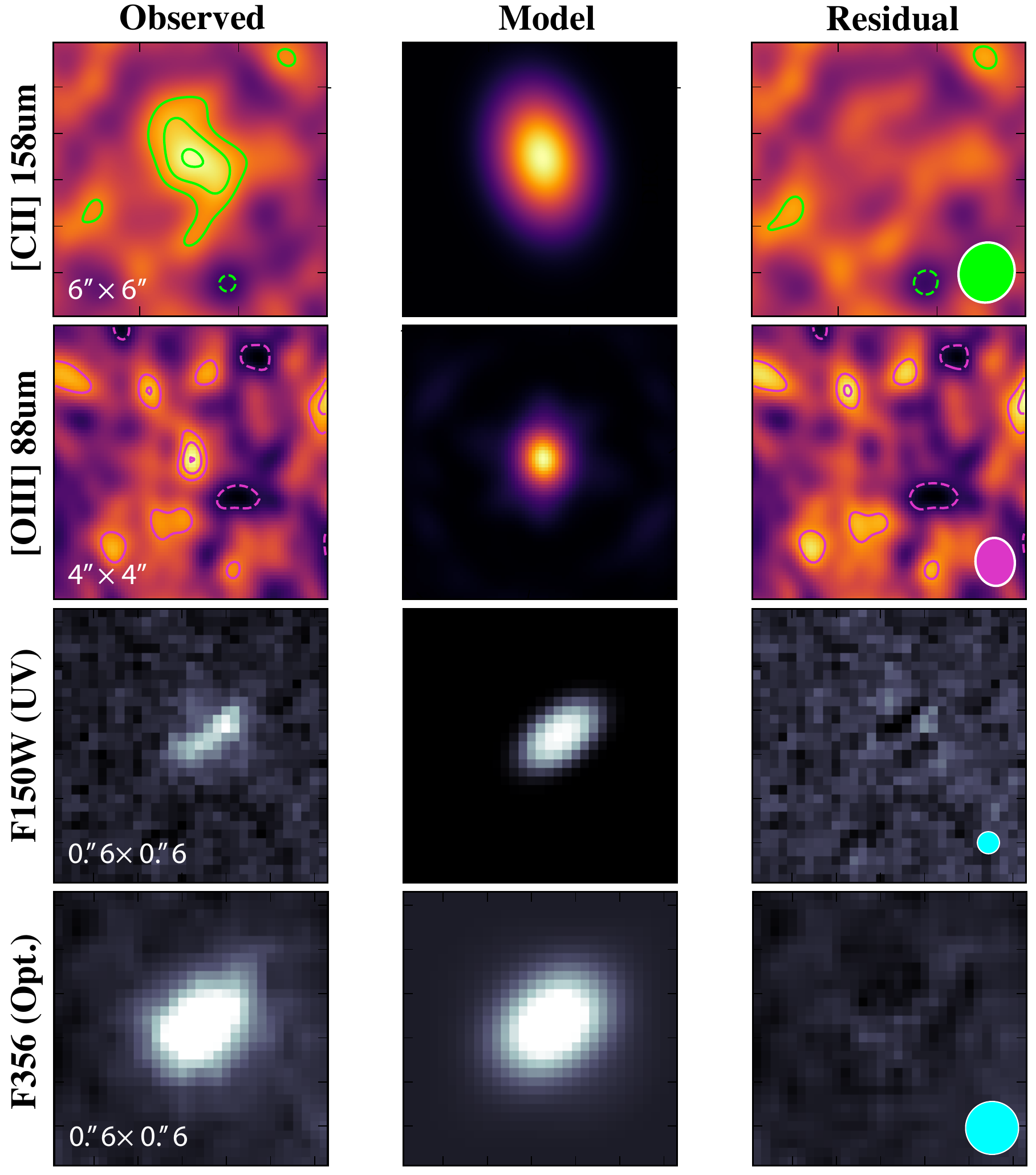}
 \caption{
Size measurement results in the image plane. 
The observed, the best-fit model, and residual (=obserevd - model) maps are shown from left to right. 
For the \cii\ and \oiii\ line with ALMA data, we use CASA {\sc imfit}, while for the rest-frame UV and optical continuum with NIRCam/F150W and F356W data, we use {\sc galfit} \citep{peng2010}. 
The ellipse (circle) at the bottom right shows the ALMA synthesized beam (NIRCam PSF). 
Note that {\sc imfit} suggests that \oiii\ is not spatially resolved, and we use the beam rescaled to the peak of the \oiii\ line as the best-fit model. 
\label{fig:size}}
\end{figure}

In Figure \ref{fig:size}, we summarize the observed, the best-fit model, and the residual maps from the best-fit size measurements for the \cii, \oiii, rest-frame UV and optical continuum emission using the ALMA and \jwst/NIRcam. The methods are described in Section \ref{sec:size}. 
We confirm that significant positive and negative pixels remain in the residual maps. 

\bibliographystyle{apj}
\bibliography{apj-jour,reference}

\end{document}